\documentstyle[aps,prb,epsfig,manuscript]{revtex}
\begin{document}
\title{Plaquette operators used in the rigorous study of ground-states of
the Periodic Anderson Model in $D = 2$ dimensions.}
\author{Zsolt~Gul\'acsi}
\address{ 
Department of Theoretical Physics, University of Debrecen, Poroszlay ut 6/C,
H-4010 Debrecen, Hungary }
\date{Dec., 2002}
\maketitle
\begin{abstract}
The derivation procedure of exact ground-states for the periodic Anderson 
model (PAM) in restricted regions of the parameter space and $D=2$ dimensions 
using plaquette operators is presented in detail. Using this procedure,
we are reporting for the first time exact ground-states for PAM in 2D
and finite value of the interaction, whose presence do not require the
next to nearest neighbor extension terms in the Hamiltonian. In order 
to do this, a completely new type of plaquette operator is introduced for PAM,
based on which a new localized phase is deduced whose physical properties are
analyzed in detail. The obtained results provide exact theoretical data which 
can be used for the understanding of system properties leading to 
metal-insulator transitions, strongly debated in recent publications in the 
frame of PAM. In the described case, the lost of the localization character
is connected to the break-down of the long-range density-density correlations
rather than Kondo physics.
\end{abstract}

\section{Introduction}

The periodic Anderson model (PAM) is one of the basic models describing 
strongly correlated systems whose characteristics fit to be interpreted in 
a two-band picture \cite{int1}. The model is largely applied in the study of 
heavy-fermion systems \cite{int2}, intermediate-valence compounds \cite{int3},
or even high critical temperature superconductors \cite{int4}. 
In contrast however with other models used in the understanding of phenomena
created by strong correlation effects, where at least in one dimension exact 
solutions exist (for example the Hubbard model \cite{int5}),
the physics described by PAM is almost exclusively interpreted based on 
approximations. This is due to the fact that only few results are exactly 
known about the behavior of PAM.

Indeed, what we know about the physical behavior of PAM in rigorous terms can 
be summarized as follows: The first results, related to the ground-state of
decorated hyper-cubic lattices in the limit of infinite interaction strength  
has been provided by Brandt and Giesekus \cite{exa1} followed by a non-magnetic
ground-state restricted to 1D and on-site repulsion $U=\infty$ case obtained
by Strack \cite{exa2}. This solution has been extended to $D=2,3$ as well, 
at $U=\infty$ \cite{exa3,exa4}. Again for infinite on-site Hubbard
repulsion has been demonstrated that at quarter filling the ground-state is 
unique with a defined total spin \cite{exa5}. It has been also underlined 
that the model becomes solvable in the case of constant and infinite range 
hopping \cite{exa6}. We further know, that for only on-site hybridization and 
without direct hopping in the correlated band: the symmetric half-filled case 
is spin-singlet \cite{exa7} and in 1D also pseudo-spin singlet
\cite{exa8}, at half filling anti-ferromagnetic correlations are present
\cite{exa9}, and in 1D,2D long-range order of ferro, anti-ferro and pairing 
type is absent \cite{exa10}. Recently have been published the first exact
ground-states at finite $U$ in 1D \cite{exa11,exa12} and 2D \cite{exa13},
respectively. Concerning 2D, the reported ground-states \cite{exa13} require  
next to nearest-neighbor (NNN) one-particle terms as well in the Hamiltonian
($\hat H$), and from the obtained solutions, especially the physical 
properties of the itinerant one has been described in detail.

The deduction of exact ground-states in $D=2$ dimensions is of large interest 
in general, and for strongly correlated systems in special. In the case of 
PAM, at finite and nonzero value of the interaction, the only one procedure 
doable working at the moment in this respect, is based on plaquette operators 
introduced in Ref.\cite{exa13}, and a decomposition of the on-site Hubbard 
interaction as described in Ref. \cite{exa11}. During this procedure, $\hat H$
is such transformed to contain in a positive semidefinite form products of 
plaquette operators. Since in 2D the 
product of plaquette operators generates NNN one-particle terms as well, the 
general impression created by the method suggests that the applicability of the
procedure is intimately connected to NNN contributions in 2D, and the deduced
exact ground-states are fingerprints of this fact.   

In this paper, presenting for the first time exact ground-states for PAM in 2D
at finite value of the interaction, in restricted regions of the 
parameter space and without NNN type of extension terms in $\hat H$, we 
demostrate that the plaquette operator procedure and the ground-states
deduced with it are not necessarily connected to the presence of NNN 
extensions in $\hat H$. In order to clarify these aspects (i) the 
plaquette operator technique is analysed in detail in general terms and 2D,
(ii) a completely new type of plaquette operator is introduced which allows
the deduction of the presented results, (iii) the obtained new localized
exact ground-state is analyzed and described in detail, and (iv) implications
of the results relating the metal-insulator transition in frame of PAM are 
presented.

The deduced new ground-state is a completely localized state. In 
order to characterize this phase, after obtaining the exact ground-state, all 
relevant ground-state expectation values and correlation functions have been 
exactly calculated and analyzed. The obtained state is paramagnetic, and based
on a coherent control which it has on the occupation number of all lattice 
sites, it introduces long-range density-density correlations into the system, 
producing a localized state.

Concerning implications to physical systems, we mention the intense activity
in the field related to the understanding of the metal-insulator transition 
(MIT) in frame of PAM. The subject has an almost 30 years of history 
\cite{bev1}, and gained renewed interest in the last period based on the
observed MIT similarities between the Hubbard model and PAM, used for example
in the understanding of the iso-structural electronically driven MIT 
transitions (like the $\gamma \to \alpha$ transition in $Ce$ compounds) 
\cite{bev2,bev3,bev4,bev5,bev6,bev7}. Since the exactly deduced ground-state
energy values presented in this paper are not containing exponential 
factors characteristique to Kondo type of behavior, the results reported here 
underline that at least in some regions of the parameter space, a 
localization-delocalization transition in frame of PAM is not necessarily 
connected to Kondo physics. 

The remaining part of the paper is structurated as follows: Section II.
presents the Hamiltonian we use, Section III. describes the plaquette operator
technique, Section IV. presents the transformation of the starting Hamiltonian 
into a new expression containing plaquette operators, Section V.  
describes the detected new exact ground-states, Section VI. concludes the 
paper, and Appendices A.- D. containing the mathematical details of the
starting points of the paper close the presentation.

\section{The expression of the Hamiltonian.}

We start with a generic PAM Hamiltonian written for 2D square lattice as
\begin{eqnarray}
\hat H = \hat T_d + \hat T_f + \hat E_f + \hat V + \hat U \: ,
\label{e8}
\end{eqnarray}
where, the contributing terms in order, are representing the kinetic energy 
of $d$ electrons ($\hat T_d$), the kinetic energy of $f$ electrons 
($\hat T_f$), the on-site $f$ electron energy ($\hat E_f$), the hybridization 
($\hat V$), and the on-site Hubbard interaction written for $f$ electrons
$\hat U = U \hat U_f$, the last contribution representing the interaction 
term, and $U > 0$ being considered during this paper. 
The presence of $\hat T_f$ in $\hat H$ is motivated by the 
overwhelming evidence, that the heavy-fermion materials contain in their real
band structure around the Fermi level ($E_F$) very narrow, hybridized bands, 
which exist already at temperatures far above the thermodynamically determined
Kondo temperature, being relatively $T$ independent and holding an accentuate
$f$ character \cite{bev8}.
 
The interaction term during this paper is exactly transformed in the form
\begin{eqnarray}
\hat U_f = \sum_i \hat n^f_{i,\uparrow} \hat n^f_{i,\downarrow} =
\hat P' + \sum_{i} ( \sum_{\sigma} \hat n^f_{i,\sigma} - 1 ) \: ,
\label{e9}
\end{eqnarray} 
where, the positive semidefinite operator 
$\hat P' = \sum_{i}( 1 - \hat n^f_{i,\uparrow} - \hat n^f_{i,\downarrow}
+ \hat n^f_{i,\uparrow} \hat n^f_{i,\downarrow})$ defined by Eq.(\ref{e9})
requires for its lowest zero eigenvalue at least one $f$ electron on every 
lattice site \cite{exa11}. As will be clarified in Section V., the 
representation presented in Eq.(\ref{e9}) is a key feature from the point of
view of the interaction term in the process of the deduction of exact 
ground-states in the frame presented here.

The hybridization $\hat V$ is considered to be 
build up from a local $\hat V_0$, and a nonlocal $\hat V_{nl}$ contribution, 
$\hat V = \hat V_0 + \hat V_{nl}$. Thus, the local one-particle terms of the 
Hamiltonian are $\hat E_f$ and $\hat V_0$ whose expressions become
\begin{eqnarray}
\hat E_f = E_f \sum_{i,\sigma} \hat n^f_{i,\sigma} \: , \quad
\hat V_0 = \sum_{i,\sigma} ( V_0 \hat d^{\dagger}_{i,\sigma} \hat f_{i,\sigma}
+ H.c. ) \: .
\label{e10}
\end{eqnarray}
The non-local one-particle contributions remain to be presented. In order to 
make the notations clear, instead of the site-numbering notation (i) we use 
here the ${\bf i}$ vectorial notation for the lattice sites. The kinetic 
energy contribution $\hat T = \hat T_d + \hat T_f$ thus is given by
\begin{eqnarray}
\hat T = \sum_{{\bf i},{\bf r},\sigma} \: ( \: t^d_{\bf r} \: 
\hat d^{\dagger}_{{\bf i},\sigma} \hat d_{{\bf i}+{\bf r},\sigma} \: + \:
t^f_{\bf r} \: \hat f^{\dagger}_{{\bf i}, \sigma} \hat f_{{\bf i}+{\bf r},
\sigma} \: + \: H.c. \: ) \: ,
\label{e11}
\end{eqnarray}
and the non-local hybridization becomes
\begin{eqnarray}
\hat V_{nl} = \sum_{{\bf i},{\bf r},\sigma} \: ( \: V^{df}_{\bf r} \:
\hat d^{\dagger}_{{\bf i}, \sigma} \hat f_{{\bf i}+{\bf r},\sigma} \: + \:
V^{fd}_{\bf r} \: \hat f^{\dagger}_{{\bf i}, \sigma} \hat d_{{\bf i}+{\bf r},
\sigma} \: + \: H.c. \: ) \: .
\label{e12}
\end{eqnarray}
The notation of the non-local hybridization matrix elements by the superscripts
$(df)$ and $(fd)$ is given by mathematical convenience, and through the 
paper 
\begin{eqnarray}
V^{df}_{\bf r} = V^{fd}_{\bf r} = V_{\bf r} \: ,
\label{e13}
\end{eqnarray}
will be considered. We note that at the level of one-particle contributions,
$\hat H$, and as a consequence $\hat T$ and $\hat V_{nl}$, contain at 
the start all contributions entering in an elementary plaquette (unit cell 
for the system) \cite{bev9}. In these circumstances, for both 
Eqs.(\ref{e11},\ref{e12}), it is important that ${\bf r}$ to be rigorously 
defined. This is because (i) we would like to represent different 
contributions correctly in term of plaquette operators, and (ii) we must 
avoid multiple counting of different terms entering in the expression of 
$\hat H$. For this reason, we mention that for a given lattice site, taking 
into account nearest neighbor (NN) and NNN contributions as well
(these will be present in an elementary plaquette), 8 hopping possibilities 
exist. From these, only 4 are taken into account 
explicitly by $\sum_{\bf r}$, and, the remaining 4 contributions are 
introduced into the Hamiltonian by the $H.c.$ operation. In these conditions, 
the defined 4 different ${\bf r}$ contributions entering in $\sum_{\bf r}$ are
\begin{eqnarray}
{\bf x} = a {\bf x_1} \: , \quad {\bf y} = a {\bf x_2} \: , \quad
{\bf x} + {\bf y} = a ({\bf x_1} + {\bf x_2}) \: , \quad
{\bf y} - {\bf x} = a ({\bf x_2} - {\bf x_1}) \: ,
\label{e14}
\end{eqnarray}
where $a$ is the lattice constant, and ${\bf x_1}$, ${\bf x_2}$, are the
versors of the $Ox$, $Oy$ axis, respectively. The hopping and hybridization 
matrix elements generated by the contributions from Eq.(\ref{e14}) are 
represented for clarity in Fig. 1. Also for clarity, the explicit expressions 
of $\hat T$ and $\hat V_{nl}$ from Eqs.(\ref{e11},\ref{e12}) are presented in 
Appendix. A.
   
In the case of concrete materials, the NNN one-particle contributions are 
small, and as a consequence are often neglected. Furthermore, it is important 
to know in rigorous terms if NNN contributions are introducing 
small corrections into the results or are able to provide qualitatively new 
effects. In the case of PAM, which is in a relatively early stage of its 
exact description, this issue must be also clarified. Because of this reason, 
during this paper, even if we start with NNN terms in $\hat H$ for technical 
reasons, we try to obtain exact ground-state solutions for PAM  in the 
absence of NNN contributions as well. This task is also enhanced by the 
aim to extend the potential possibilities of the 
plaquette operator procedure we use. Starting from these motivations, we are 
reporting in this paper for the first time exact ground-states for PAM in 2D 
at finite and nonzero $U$, in the absence of NNN extension terms in $\hat H$.

In order to be able to obtain exact ground-states in $D = 2$ dimensions for
the PAM Hamiltonian presented in Eq.(\ref{e8}), we use a plaquette operator
procedure which will be described in details in the following Section. 
Concerning the method itself, in our knowledge, it is used now for the second 
time (see also \cite{exa13}), and other methods in obtaining exact 
ground-states for PAM in $D = 2$ dimensions are not known at the moment. In 
principle, the procedure can be applied for other model Hamiltonians as well 
containing itinerant degrees of freedom. The technique needs the 
transformation of $\hat H$ in a positive semidefinite expression based on
plaquette operators. In 2D, the plaquette operators (as block operator units 
used for description) generate local, NN and NNN terms as well. This suggests 
that also the studied 2D Hamiltonian must contain such type of contributions. 
If this would be the case, the application possibility of the procedure in 2D
would be strongly limited to Hamiltonians that contain NNN extension terms as 
well. We demonstrate in this paper that this impression is not correct, and 
the procedure can be extended and applied even in the absence of NNN 
contributions in $\hat H$. We further mention that for Hamiltonians 
containing main long range terms (next to NNN or higher range contributions), 
the block unit used for description must be enlarged. 

\section{Plaquette operators used for the transformation of the Hamiltonian.}

Let us consider a 2D finite $N_{\Lambda} = N_{L} \times N_{L}$ square lattice, 
with lattice constant $a$. In order to identify the lattice sites, we are 
numbering them by $i$ starting from the left-down corner, taking into account 
first the lowest row, and inside a row counting from left to right 
(we mention that for a vectorial position notation we are going to use
${\bf i}$ instead of i, when this is necessary). For example, in
the simple case of $N_{L} = 4$, we obtain the lattice site
numbering presented in Fig. 2. As can be seen  in this figure, we are denoting
by $Pi$ the elementary plaquettes. Using this notation, we start from
the lowest elementary plaquette row, counting from left to right inside a row,
and then going upward with the notation.

Taking periodic boundary conditions into account in both directions, the 
number of plaquettes becomes equal to the number of lattice sites 
$N_{\Lambda}$. In this case, it is advantageous to denote every plaquette $Pi$
by its down-left corner $j$, as $p_j$. Concerning the notation of a
plaquette through its down-left corner, for clarity we mention that for 
example, in Fig. 2., the plaquette $P5$ defined by the lattice sites 
$(6, 7, 10, 11)$ becomes $p_6$, or, the plaquette $P7$, defined by the 
lattice sites $(9, 10, 13, 14)$ becomes $p_9$, etc.

Let us now consider for pedagogical reasons, some 
$\hat c^{\dagger}_i$ fermionic operators creating particles on lattice sites 
within the system. In general, the $\hat c_i$ operators can be labelled also 
by a supplementary $\alpha$ index which contains all relevant quantum numbers 
as well (in the case of PAM, $\alpha = (\sigma, g)$, where $\sigma = \uparrow,
\downarrow$ denotes the spin, and $g=d,f$ the type of particle). In this 
Section, being interested in the presentation of the method, we are neglecting
the $\alpha$ index for simplicity. If the reader understands how the procedure
works, the presented relations can be easily generalized for $\hat c_{i,
\alpha}$ as well. 
 
Using the $\hat c_i$ operators, plaquette operators can be 
constructed by a linear combination of $\hat c_i$ acting on the 
corners of an elementary plaquette. We are denoting the coefficients of this 
linear combination by $a_{p,i,c}$, where $p$ denotes the plaquette, $i$ 
labels a given corner of the plaquette $p$ analyzed, and $c$ denotes the type 
of operator considered (this becomes the $\alpha$ index when $\hat 
c_{i,\alpha}$ is used instead of $\hat c_i$), respectively. For example, 
in case of the plaquettes $p(i1)$ and $p(i1+1)$ from Fig. 3A. we obtain
\begin{eqnarray}
&&\hat A_{i1} \: = \: a_{p(i1),i1,c} \hat c_{i1} + a_{p(i1),i1+1,c} 
\hat c_{i1+1} + a_{p(i1),j1,c} \hat c_{j1} + a_{p(i1),j1+1,c} 
\hat c_{j1+1} \: ,
\nonumber\\
&&\hat A_{i1+1} = a_{p(i1+1),i1+1,c} \hat c_{i1+1} + a_{p(i1+1),i1+2,c} 
\hat c_{i1+2} + a_{p(i1+1),j1+1,c} \hat c_{j1+1} + a_{p(i1+1),j1+2,c} 
\hat c_{j1+2} \: .
\label{e1}
\end{eqnarray}
Working with plaquettes, we must observe that all one-particle 
contributions of a given Hamiltonian  can be obtained starting from 
plaquette operators. 
For example, let us consider the hopping matrix element connecting 
the nearest-neighbor lattice sites $(i1+1, j1+1)$ from Fig. 3A, namely
$\hat T_{i1+1,j1+1} = ( t^c_{i1+1,j1+1} \hat c^{\dagger}_{i1+1} \hat c_{j1+1} 
+ H.c.)$. This Hamiltonian contribution can be obtained for example, from the 
expression $\hat A_{i1}^{\dagger} \hat A_{i1} + \hat A_{i1+1}^{\dagger} 
\hat A_{i1+1}$. Indeed, we have
\begin{eqnarray}
\hat A_{i1}^{\dagger} \hat A_{i1} + \hat A_{i1+1}^{\dagger} \hat A_{i1+1} =
\hat T_{i1+1,j1+1} + \hat O \: ,
\label{e2}
\end{eqnarray}
where, the operator $\hat O$ concentrates all the other terms obtained from 
the left side of Eq.(\ref{e2}). The operator $\hat O$ will not be neglected  
in our considerations. It contains 30 terms that can be easily calculated from 
Eq.(\ref{e1}) (see also Appendix B.). The important aspect here, which must be
keeped in mind, is that $\hat O$ do not contains contributions entering in 
$\hat T_{i1+1,j1+1}$. Otherwise, the concrete expression of $\hat O$ is not 
important at the moment. The relation from Eq.(\ref{e2}) is obtained
since, $\hat A_{i1}^{\dagger} \hat A_{i1}$ gives rise 
to $(a^{*}_{p(i1),i1+1,c} a_{p(i1),j1+1,c} \hat c^{\dagger}_{i1+1} 
\hat c_{j1+1} + H.c.)$, and $\hat A_{i1+1}^{\dagger} \hat A_{i1+1}$ creates 
the term
$(a^{*}_{p(i1+1),i1+1,c} a_{p(i1+1),j1+1,c} \hat c^{\dagger}_{i1+1}
\hat c_{j1+1} + H.c.)$, respectively. Because the bond $(i1+1,j1+1)$ is not 
present in other elementary plaquettes, even if we take into consideration all
the plaquettes from the whole lattice in a sum of the form 
$\sum_i \hat A^{\dagger}_i \hat A_i$, the hopping matrix element 
$t^c_{i1+1,j1+1}$ becomes unambiguously expressed as
\begin{eqnarray}
t^c_{i1+1,j1+1} = a^{*}_{p(i1),i1+1,c} a_{p(i1),j1+1,c} +
a^{*}_{p(i1+1),i1+1,c} a_{p(i1+1),j1+1,c} \: .
\label{e3}
\end{eqnarray}
The obtained Eq.(\ref{e3}) shows that the Hamiltonian parameters (at least the
one-particle once in the present case), can be expressed in term of plaquette
operator parameters if we succeed to express the corresponding Hamiltonian 
terms into a sum of the form $\sum_i \hat A^{\dagger}_i \hat A_i$. 

Similarly, the next-nearest-neighbor hopping amplitude for the $(i1,j1+1)$ 
hopping from the plaquette $p(i1)$ of Fig. 3A., contained in the Hamiltonian
term $\hat T_{i1,j1+1} = ( t^c_{i1,j1+1} \hat c^{\dagger}_{i1} \hat c_{j1+1} +
H.c. )$ becomes
\begin{eqnarray}
t^c_{i1,j1+1} = a^{*}_{p(i1),i1,c} a_{p(i1),j1+1,c} \: .
\label{e3a}
\end{eqnarray}
In Eq.(\ref{e3a}) only the plaquette operator product $\hat A^{\dagger}_{i1}
\hat A_{i1}$ contributes, because, the NNN hopping described
by $\hat T_{i1,j1+1}$ is contained only in the plaquette $p(i1)$. These 
examples illustrate that plaquette operators can be extremely useful
in the study of different model Hamiltonians $\hat H$, since as seen from 
Eq.(\ref{e2}), different emerging contributions in $\hat H$ can be represented
in diagonal, or positive semidefinite form via the operators $\hat A_i$. 
As can be observed from Eqs.(\ref{e3},\ref{e3a}), a such type of 
representation in term of plaquette operators, from the point of
view of $\hat H$ parameters, simply means a parametrization in term of
plaquette operator coefficients $a_{p,i,c}$. For this to be possible, the
plaquette operator products summed up over lattice sites of the form 
$\sum_i \hat A^{\dagger}_i \hat A_i$ (i) must generate terms present in 
$\hat H$, or (ii) must generate terms that are constants of motion (for example
total number of particles, or lattice sites), or (iii) must generate terms 
that can be cancelled out if the (i) and (ii) conditions cannot be applied. 
We will return back to this problem after presenting the new plaquette 
operators defined in this paper (see after Eq.(\ref{e7})), 
and the following Section exemplifies in detail a such type of transformation.

When the one-particle $\hat H$ parameters are not local (for example
$t^c_{i1,j1} = t^c_{i1+1,j1+1} = t^c_{i1+2,j1+2}$ for all vertical 
nearest-neighbor
hoppings), which means
\begin{eqnarray}
t^c_{i1+1,j1+1} = t_y^c \: ,
\label{e4}
\end{eqnarray}
the parameters $a_{p,i,c}$ of different plaquette operators are not 
independent. In the case of translational
invariant Hamiltonians, we can chose for example translational invariant 
plaquette operator parameters as illustrated by Fig. 3B. Denoting the sites
inside a given plaquette starting from the down-left corner and counting 
anti-clockwise, the corners of the plaquette $p(i1)$ (and $p(i1+1)$) in 
Fig.3B., will be denoted by $n$ (and $m$), respectively. Given by the 
considered translational invariance of plaquette operators, the plaquette 
operator parameters of the plaquettes $p(i1)$ and $p(i1+1)$ with $n = m = 
\tau$ equal indices will have the same value $a_{\tau,c}$, $\tau = 1,2,3,4$. 
This property is extended as well to all plaquettes. In the examples contained
in Fig. 3B., the plaquette operators $\hat A_{i1}$, $\hat A_{i1+1}$ become in 
this case
\begin{eqnarray}
&&\hat A_{i1} = a_{1,c} \hat c_{i1} + a_{2,c} \hat c_{i1+1} +
a_{4,c} \hat c_{j1} + a_{3,c} \hat c_{j1+1} \: ,
\nonumber\\
&&\hat A_{i1+1} = a_{1,c} \hat c_{i1+1} + a_{2,c} 
\hat c_{i1+2} + a_{4,c} \hat c_{j1+1} + a_{3,c} 
\hat c_{j1+2} \: .
\label{e5}
\end{eqnarray}   
From Eqs.(\ref{e3},\ref{e4},\ref{e5}), the unique NN hopping 
matrix element in $y$ direction of $c$ particles, based on Eq.(\ref{e3}) 
becomes
\begin{eqnarray}
t^c_y = a^{*}_{2,c} a_{3,c} + a^{*}_{1,c} a_{4,c} \: ,
\label{e6}
\end{eqnarray}
and, from Eqs.(\ref{e3a},\ref{e5}), the unique NNN hopping of the same 
particles along the main diagonal of every elementary plaquette 
will be described by
\begin{eqnarray}
t^c_{x+y} = a^{*}_{1,c} a_{3,c} \: .
\label{e6a}
\end{eqnarray}  
Similarly, all one-particle Hamiltonian matrix elements can be expressed in
term of plaquette operator parameters. When the so obtained equations (as
Eqs.(\ref{e6},\ref{e6a})) allow
solutions for the plaquette operator parameters (this is possible usually in a
restricted parameter space region ${\cal{P}}_H$ determined by the values of
$\hat H$ parameters), the one-particle part of the
Hamiltonian can be expressed via $\sum_i \hat A_{i}^{\dagger} \hat A_{i}$
(see Eq.(\ref{e2})). Based on these relations and using for example 
properties related to positive semidefinite operators, the Hamiltonian of the 
system can be diagonalized exactly, at least for the ground-state, inside 
${\cal{P}}_H$.

After testing this method in 1D \cite{exa11,exa12} (using bonds instead of 
plaquettes), a such type of procedure has been recently used by us 
\cite{exa13} in order to provide the first exact ground-state wave-functions 
for the periodic Anderson model in 2D in restricted regions of the parameter 
space. This has been done by choosing
$\hat c_i = \hat d_{i,\sigma}, \hat f_{i,\sigma}$ for $d,f$ electrons 
with ${\it fixed}$ ${\it spin}$ in PAM, and defining based on this choice, 
the $\hat A_{i,\sigma}$ spin-dependent plaquette operators containing 
spin-independent $a_{n,g}$ parameters with $g=d,f$, $n=1,2,3,4$ as follows
(the example is taken for the plaquette $p(i1)$ of Fig. 3B.)
\begin{eqnarray}
&&\hat A_{i1,\sigma} = a_{1,d} \hat d_{i1,\sigma} + a_{1,f} \hat f_{i1,\sigma}
+ a_{2,d} \hat d_{i1+1,\sigma} + a_{2,f} \hat f_{i1+1,\sigma} +
\nonumber\\
&&a_{3,d} \hat d_{j1+1,\sigma} + a_{3,f} \hat f_{j1+1,\sigma}
+ a_{4,d} \hat d_{j1,\sigma} + a_{4,f} \hat f_{j1,\sigma} \: .
\label{e6b}
\end{eqnarray}
 
The obtained ground-state solutions based on Eq.(\ref{e6b}) were connected to
$3/4$ filling \cite{exa13}, and are highly non-trivial states. One of them is a
completely localized state, and the second one is itinerant, with the momentum
distribution function for the half-filled upper diagonalized band as shown in 
Fig. 4, presenting a clear evidence of (exactly deduced) non-Fermi liquid 
behavior in normal phase and $D=2$ spatial dimensions. This shows that the
procedure detects ground-states which are far to be trivial. However, the 
inconvenience of the plaquette operator from Eq.(\ref{e6b}) is that via 
$\sum_{i,\sigma} \hat A^{\dagger}_{i,\sigma} \hat A_{i,\sigma}$ it creates 
NNN terms, these must be present in $\hat H$ as well, so the deduced 
ground-states, and the procedure itself, seem to be related to the presence 
of NNN extensions in the Hamiltonian. We present below how this inconvenience 
can be removed.  

For this reason, we must observe, that the choice of the operators 
$\hat c_i$ in Eq.(\ref{e5}) and the form of the plaquette operator itself is 
not fixed a ${\it priori}$. This means that the possibility
presented in Eq.(\ref{e6b}) for the plaquette operators is not unique, even if
we are interested in the study of a fixed model (as PAM in the present case).
As a consequence, we can chose other possible forms for the plaquette 
operators, and using them, we can deduce other ground-states in other regions 
of the $T=0$ phase diagram of the model. To exemplify this statement,
in the present paper we define for the decomposition of the studied PAM 
Hamiltonian in translational invariant case, a completely new type of 
plaquette operators $\hat A_{\bf i}$ and $\hat B_{\bf i}$. Each of these has 
different plaquette operator parameters $a_{n,g,\sigma}$ and $b_{n,g,\sigma}$, 
$n=1,2,3,4$, $g=d,f$, $\sigma= \uparrow,\downarrow$. Furthermore, both 
plaquette operators $\hat A_{\bf i}$ and $\hat B_{\bf i}$ are containing both 
spin components with different numerical prefactors, i.e. $a_{n,g,\sigma}$, 
$b_{n,g,\sigma}$ are considered independent, and $\sigma$ dependent. 
Exemplifying the new form for the case of the plaquette $p(i1)$ of Fig. 3B.,
where ${\bf i}$ denotes the vectorial position of the site $i1$, the 
new $\hat A_{\bf i}$ operator is defined as
\begin{eqnarray}
\hat A_{i1} &=&
a_{1,d,\uparrow} \hat d_{i1,\uparrow} + 
a_{2,d,\uparrow} \hat d_{i1+1,\uparrow} +
a_{3,d,\uparrow} \hat d_{j1+1,\uparrow} +
a_{4,d,\uparrow} \hat d_{j1,\uparrow} 
\nonumber\\
&+& a_{1,d,\downarrow} \hat d_{i1,\downarrow} + 
a_{2,d,\downarrow} \hat d_{i1+1,\downarrow} +
a_{3,d,\downarrow} \hat d_{j1+1,\downarrow} +
a_{4,d,\downarrow} \hat d_{j1,\downarrow} 
\nonumber\\
&+&
a_{1,f,\uparrow} \hat f_{i1,\uparrow} + 
a_{2,f,\uparrow} \hat f_{i1+1,\uparrow} +
a_{3,f,\uparrow} \hat f_{j1+1,\uparrow} +
a_{4,f,\uparrow} \hat f_{j1,\uparrow} 
\nonumber\\
&+& a_{1,f,\downarrow} \hat f_{i1,\downarrow} + 
a_{2,f,\downarrow} \hat f_{i1+1,\downarrow} +
a_{3,f,\downarrow} \hat f_{j1+1,\downarrow} +
a_{4,f,\downarrow} \hat f_{j1,\downarrow} \: . 
\label{e7}
\end{eqnarray}
Similar expression is used for the $\hat B_{\bf i}$ operator as well for the 
same plaquette $p(i1)$, in which, the plaquette operator parameters are 
considered $b_{n,g,\sigma}$, instead of $a_{n,g,n}$. Note the plaquette 
independent values of the $a_{n,g,\sigma}$ and $b_{n,g,\sigma}$ 
parameters, which, as explained in this Section, is given by the translational
invariance of the considered system. 

Comparing Eq.(\ref{e6b}) and Eq.(\ref{e7}), we realize that the
$\hat A_{i,\sigma}$ plaquette operators for both $\sigma=\uparrow,\downarrow$
values have 8 independent $a_{n,g}$ parameters, while in the present case, for
both $\hat A_{i}$, $\hat B_{i}$ operators, the number of independent plaquette
operator parameters is 32. This enlargement of the number of parameters give us
the possibility to demonstrate that the described procedure is able to detect
also ground-states whose presence do not require the NNN
terms in $\hat H$ of the system, even if the $\hat A^{\dagger}_{\bf i}
\hat A_{\bf i}$ products are providing such type of terms at the start. The key
feature for this to work is the presence of two plaquette operators $\hat A_i$
and $\hat B_i$ containing different spin-dependent coefficients. Indeed, in 
this case, by $\sum_i \hat B^{\dagger}_i \hat B_i$ we can cancel out  
not only the $\uparrow \downarrow$ terms created by $\sum_i \hat A^{\dagger}_i
\hat A_i$ which are not present in $\hat H$ (these would represent for example
hopping terms containing spin-flip), but also the NNN terms
generated by $\sum_i \hat A^{\dagger}_i \hat A_i$. Because of this reason 
becomes to be possible to obtain the expression of a Hamiltonian not 
containing NNN contributions in term of plaquette operator products which 
create such type of elements. The concrete transformation of the Hamiltonian is
presented in the following Section.

\section{The Hamiltonian written in term of plaquette operators.}

Comparing the expression of the Hamiltonian presented in the previous Section
together with the explicitations contained in Appendices A. and B., Eqs.(
\ref{e10},\ref{a1},\ref{a2},\ref{b1}), we realize that the following relation 
holds
\begin{eqnarray}
&&\hat T_d + \hat T_f + \hat V_0 + \hat V_{nl} =
- \sum_{\bf i}^{N_{\Lambda}} \hat A^{\dagger}_{\bf i} \hat A_{\bf i} 
- \sum_{\bf i}^{N_{\Lambda}} \hat B^{\dagger}_{\bf i} \hat B_{\bf i} 
\nonumber\\
&&+ \sum_{\sigma} [\sum_{n=1}^4(|a_{n,d,\sigma}|^2 + |b_{n,d,\sigma}|^2)]
\sum_{\bf i}^{N_{\Lambda}} \hat d^{\dagger}_{{\bf i},\sigma} \hat d_{{\bf i},
\sigma} +
\sum_{\sigma} [\sum_{n=1}^4(|a_{n,f,\sigma}|^2 + |b_{n,f,\sigma}|^2)]
\sum_{\bf i}^{N_{\Lambda}} \hat f^{\dagger}_{{\bf i},\sigma} \hat f_{{\bf i},
\sigma} \: ,
\label{e15}
\end{eqnarray}
if, the hopping and hybridization  matrix elements are related to the 
parameters of the plaquette operators $\hat A_{\bf i}$ and $\hat B_{\bf i}$
via
\begin{eqnarray}
F(t^{g}_{\bf r},V_{\bf r}; a_{n,g,\sigma}, b_{n,g,\sigma}) = 0 \: ,
\label{e16}
\end{eqnarray}
where the non-linear system of equations from Eq.(\ref{e16}) is presented
explicitly in the Appendix. C., and $g = d,f$. These equations arise 
as Eqs.(\ref{e6},\ref{e6a}) in Section I. The system of equations
Eq.(\ref{e16}) must be considered as containing known $\hat H$ parameters 
($t^{g}_{\bf r}, V_{\bf r}$), and unknown plaquette operator numerical
prefactors ($a_{n,g,\sigma}, b_{n,g,\sigma}$). In fact, a simple (but lengthy)
algebraic calculation shows that Eq.(\ref{e15}) exactly holds if the relations
between the parameters of $\hat H$ and the numerical prefactors of the 
plaquette operators, presented explicitly in Appendix C, are satisfied.
The number of equations contained in Eq.(\ref{e16}) is
$70$, and the 32 unknown complex plaquette operator parameters provides 64
unknown variables (the real and imaginary parts). These are entering in 
Eq.(\ref{e16}) in a nonlinear, but complex-algebraic manner. Since the number 
of equations is higher that the number of unknown variables, solutions will be 
allowed only if some inter-dependences (fixed by Eq.(\ref{e16})) will be 
present between the $\hat H$ parameters. These relations contribute
in the definition of ${\cal P}_H$ (see also the observations below Eq.(
\ref{e24})).

We underline that since the structure of plaquette operators used in this paper
(see Eq.(\ref{e7})) is completely different from the structure of the plaquette
operators from Eq.(\ref{e6b}) (in that case, instead of Eq.(\ref{e16}), 
we have 17 equations presented in Eq.(9) of Ref.\cite{exa13}, containing 16 
unknown variables), the problem set up here, from mathematical point of view,
is completely different from that analysed in our previous work.

We also note that as can be seen from Appendix B, $\sum_{\bf i} \hat A_{
\bf i}^{\dagger} \hat A_{\bf i}$ introduces $(\uparrow,\downarrow)$ like terms
as well, which are missing from the Hamiltonian. Because of this reason we need
a second plaquette operator product $\sum_{\bf i} \hat B^{\dagger}_{\bf i}
\hat B_{\bf i}$, whose role is exactly to cancel out these supplementary 
terms not present in $\hat H$, Eq.(\ref{e8}). Furthermore, the presence of
$\sum_{\bf i} \hat B^{\dagger}_{\bf i} \hat B_{\bf i}$ allows also to cancel 
out the NNN terms created by $\sum_{\bf i} \hat A^{\dagger}_{\bf i} 
\hat A_{\bf i}$. Via Eq.(\ref{e15}), this give as the possibility to express
$\hat H$ through $\sum_{\bf i} ( \hat A^{\dagger}_{\bf i} \hat A_{\bf i} +
\hat B^{\dagger}_{\bf i} \hat B_{\bf i} )$ even in the absence of NNN terms in
the Hamiltonian, and to obtain ground-state wave-functions in this case as 
well.

Using now Eqs.(\ref{e9},\ref{e10}), we have
\begin{eqnarray}
\hat U + \hat E_f = U \hat P' + (E_f + U) \sum_{{\bf i},\sigma}
\hat f^{\dagger}_{{\bf i},\sigma} \hat f_{{\bf i},\sigma} - U N_{\Lambda} \: ,
\label{e17}
\end{eqnarray}
where, the positive semidefinite operator $\hat P'$ has been defined in Sec.II.
Adding Eq.(\ref{e17}) to Eq.(\ref{e15}) and using for the plaquette operators
the anti-commutation property presented in Eq.(\ref{b2}), we find
\begin{eqnarray}
\hat H &=& \sum_{\bf i} ( \hat A_{\bf i} \hat A^{\dagger}_{\bf i} +
\hat B_{\bf i} \hat B^{\dagger}_{\bf i} ) + U \hat P' - N_{\Lambda}
(U + K^d_{\uparrow} + K^d_{\downarrow} + K^f_{\uparrow} + K^f_{\downarrow} )
\nonumber\\
&+& 
\hat N^d_{\uparrow} K^d_{\uparrow} + \hat N^d_{\downarrow} K^d_{\downarrow} +
\hat N^f_{\uparrow} K^f_{\uparrow} + \hat N^f_{\downarrow} K^f_{\downarrow} +
(E_f + U) (\hat N^f_{\uparrow} + \hat N^f_{\downarrow}) \: ,
\label{e19}
\end{eqnarray}
where, the introduced constants are defined by
$K^{g}_{\sigma} = \sum_{n=1}^4 (|a_{n,g,\sigma}|^2 + |b_{n,g,\sigma}
|^2)$, and the particle number operators by 
$\hat N^{g}_{\sigma} = \sum_{\bf i} \hat n^{g}_{\sigma}$, with $g=d,f$. 
Imposing the relations
\begin{eqnarray}
K^d_{\uparrow} = K^d_{\downarrow} = K \: , \quad
K^f_{\uparrow} = K^f_{\downarrow} = K^f \: , \quad E_f + U = K - K^f \: ,
\label{e20}
\end{eqnarray}
the expression of $\hat H$ from Eq.(\ref{e19}) becomes
\begin{eqnarray}
\hat H &=& \sum_{\bf i} ( \hat A_{\bf i} \hat A^{\dagger}_{\bf i} +
\hat B_{\bf i} \hat B^{\dagger}_{\bf i} ) + U \hat P' + K \hat N
- N_{\Lambda} (4K - 2E_f - U) \: .
\label{e21}
\end{eqnarray}
Since we are working at fixed number of particles $N$, from Eq.(\ref{e21})
we obtain
\begin{eqnarray}
\hat H = \hat P + E_g \: ,
\label{e22}
\end{eqnarray}
where $\hat P$ for $U > 0$ is a positive semidefinite operator defined by
\begin{eqnarray}
\hat P = \sum_{\bf i} ( \hat A_{\bf i} \hat A^{\dagger}_{\bf i} +
\hat B_{\bf i} \hat B^{\dagger}_{\bf i} ) + U \hat P' \: ,
\label{e23}
\end{eqnarray}
and the number $E_g$ is given by
\begin{eqnarray}
E_g = K N - N_{\Lambda} (4K -2 E_f - U) \: .
\label{e24}
\end{eqnarray}
The transformation of Eq.(\ref{e8}) into Eq.(\ref{e22}) is possible only if 
the system of equations Eqs.(\ref{e16},\ref{e20}) allows solutions for the
plaquette operator parameters. The presence of these solutions will be 
possible only on restricted domains ${\cal{P}}_H$ of the parameter space of 
the problem given by the inter-dependences between the $\hat H$ parameters  
mentioned below Eq.(\ref{e16}). As a consequence, the solutions that will be 
presented below are valid only in this ${\cal{P}}_H$ region.

\section{Exact ground-state wave-function solutions.}

In this Section we are presenting first the derivation of the exact 
ground-states, then we discuss the possible solutions for the plaquette 
operator parameters, and finally, we analyse in extreme details the solution 
obtained for zero NNN contributions.
 
\subsection{The derivation of the exact ground-states.}

Starting from Eq.(\ref{e22}), taking into account that $\hat P$ is a positive
semidefinite operator, we realize that the ground-state of 
$\hat H = \hat P + E_g$ is the wave function $|\Psi_g\rangle$, for which
$\hat P |\Psi_g\rangle = 0$. To find $|\Psi_g\rangle$, we have to keep in mind
Eq.(\ref{e23}) which defines $\hat P$. Given by
\begin{eqnarray}
\hat A_{\bf i}^{\dagger} \hat A_{\bf i}^{\dagger} = 0 \: ,
\hat B_{\bf i}^{\dagger} \hat B_{\bf i}^{\dagger} = 0 \: ,
\hat A_{\bf i}^{\dagger} \hat A_{\bf j}^{\dagger} = 
- \hat A_{\bf j}^{\dagger} \hat A_{\bf i}^{\dagger} \: ,
\hat B_{\bf i}^{\dagger} \hat B_{\bf j}^{\dagger} = 
- \hat B_{\bf j}^{\dagger} \hat B_{\bf i}^{\dagger} \: ,
\hat A_{\bf i}^{\dagger} \hat B_{\bf j}^{\dagger} = 
- \hat B_{\bf j}^{\dagger} \hat A_{\bf i}^{\dagger} \: ,
\label{e25}
\end{eqnarray}
we observe that the plaquette operator part of Eq.(\ref{e23}) applied to
$\prod_{\bf i} \hat A^{\dagger}_{\bf i} \hat B^{\dagger}_{\bf i}$ gives zero.
Furthermore, since $\hat P'$ requires for its zero (and minimum) eigenvalue
at least one $f$-electron on every lattice site, we add to the ground-state 
the contribution $ \hat F_{\mu} = \prod_{\bf i} 
(\mu_{{\bf i},\uparrow} \hat f^{\dagger}_{{\bf i}, \uparrow} + 
\mu_{{\bf i},\downarrow} \hat f^{\dagger}_{{\bf i}, \downarrow})$, where 
$\mu_{{\bf i},\sigma}$ are arbitrary coefficients. As a consequence, the 
ground-state with the property $\hat P |\Psi_g\rangle = 0$ becomes
\begin{eqnarray}
|\Psi_g \rangle = \prod_{\bf i} [\hat A^{\dagger}_{\bf i} \hat B^{\dagger}_{
\bf i} (\mu_{{\bf i},\uparrow} \hat f^{\dagger}_{{\bf i}, \uparrow} + 
\mu_{{\bf i},\downarrow} \hat f^{\dagger}_{{\bf i}, \downarrow})] |0 \rangle
\: ,
\label{e26}
\end{eqnarray}
where, $|0\rangle$ is the bare vacuum with no fermions present. The product in
Eq.(\ref{e26}) must be taken over all lattice sites. Because of this reason,
the product of the creation operators in Eq.(\ref{e26}) introduces $N = 3 
N_{\Lambda}$ particles within the system, so the deduced ground-state 
wave-function corresponds to $3/4$ filling. All degeneration possibilities of 
the ground-state are contained in Eq.(\ref{e26}), since the wave function
with the property $\hat P |\Psi\rangle = 0$ at $3/4$ filling always can be 
written in the presented $|\Psi_g\rangle$ form. We underline however that PAM
contains two hybridized bands, and $3/4$ filling for a two-band system means 
in fact half filled upper hybridized band (the lower band being completely 
filled up). 

The wave-vector $|\Psi_g\rangle$
represents the ground-state of the starting Hamiltonian, only if Eq.(\ref{e8})
can be transformed in Eq.(\ref{e22}). This is possible only if we are situated
inside the region ${\cal{P}}_H$ of the parameter space, i.e. the system of 
equations Eq.(\ref{e16}) detailed in Appendix C. allows solutions for the
plaquette operator parameters, in conditions in which also the constrains
from Eqs.(\ref{e13},\ref{e20}) hold. In the remaining part of the paper we will
concentrate on these possible solutions. 

We underline, that $|\Psi_g\rangle$ presented in Eq.(\ref{e26}) describes 
rigorously only the $U > 0$ case, since the presence of the $\hat F_{\mu}$ 
operator into the ground-state is required only by the non-zero $U$ value.
As a consequence, the ground-state at $U=0$ cannot be expressed in the
form presented in Eq.(\ref{e26}). 
We emphasize that the differences between Eq.(\ref{e26}) and the ground-states
deduced previously \cite{exa13} are present because instead of 
$\prod_{{\bf i},\sigma}\hat A^{\dagger}_{{\bf i},\sigma}$ obtained in the old
case with $\hat A_{{\bf i},\sigma}$ defined by Eq.(\ref{e6b}), we now have in 
the ground-state wave function $\prod_{\bf i} \hat A^{\dagger}_{\bf i}
\hat B^{\dagger}_{\bf i}$.

Before going further, we mention that the physical properties of the 
ground-state wave-function written mathematically in Eq.(\ref{e26}) strongly
depend on the nature of the concrete solutions provided by Eq.(\ref{e16}).

\subsection{Solutions for the plaquette operator parameters.}

The solutions for the plaquette operator parameters which lead to the 
ground-state $|\Psi_g\rangle$ must be obtained solving together
Eqs.(\ref{e13},\ref{e20},\ref{c1}). These taken together represent 74 
nonlinear complex-algebraic coupled equations, so a relatively difficult 
mathematical task.

A study of the next-nearest neighbor contributions entering in Eq.(\ref{c1}) 
shows that the solutions exist only if the following inter-dependences are
present between the plaquette operator parameters
\begin{eqnarray}
&&\frac{a^{*}_{1,d,\sigma}}{b^{*}_{1,d,\sigma}} = -
\frac{b_{3,d,-\sigma}}{a_{3,d,-\sigma}} = -
\frac{b_{3,f,-\sigma}}{a_{3,f,-\sigma}} = 
\frac{a^{*}_{1,f,\sigma}}{b^{*}_{1,f,\sigma}} = x_{\sigma} \: ,
\nonumber\\
&&\frac{a^{*}_{2,d,\sigma}}{b^{*}_{2,d,\sigma}} = -
\frac{b_{4,d,-\sigma}}{a_{4,d,-\sigma}} = -
\frac{b_{4,f,-\sigma}}{a_{4,f,-\sigma}} = 
\frac{a^{*}_{2,f,\sigma}}{b^{*}_{2,f,\sigma}} = y_{\sigma} \: ,
\label{e27}
\end{eqnarray}
where, $x_{\sigma}, \: y_{\sigma}$ are complex, finite, nonzero, otherwise
arbitrary parameters defined by the ratios presented in Eq.(\ref{e27}). Using
Eq.(\ref{e27}), the studied system of equations can be completely transcribed
for the $b_{n,g,\sigma}$ unknown variables with $n=1,2,3,4$; $g=f,d$;
$\sigma=\uparrow,\downarrow$ (the $a_{n,g,\sigma}$ parameters being given
through $b_{n,g,\sigma}$ via Eq.(\ref{e27})). Since the so obtained equations
for the $b_{n,g,\sigma}$ variables are representing the starting point of the
description of physical properties provided by the deduced ground-states, 
they are presented in Eqs.(\ref{d1},\ref{d2}) of Appendix D. Starting from 
this moment, we must solve the system of equations presented in Appendix D.

We have found for the system of equations Eqs.(\ref{d1},\ref{d2}) several 
mathematical solutions, which will be briefly presented below.

a) Taking $x_{\uparrow} = y_{\uparrow} = y, \: x_{\downarrow}= y_{\downarrow}=
- 1/y^{*}$, we find the first class of solutions. The interesting aspect of 
this case is that the 20 equations contained in Eq.(\ref{d1}) are automatically
satisfied, and we must concentrate only on equations presented in 
Eq.(\ref{d2}). This last system provides a solution for $b_{n,g,\downarrow} =
y b_{n,g,\uparrow}$, $g = d,f$, which however do not has new aspects in 
comparison to the solutions we find in Ref.(\cite{exa13}).

b) As can be seen, in order to obtain  new solutions, we must take
$x_{\uparrow} \ne y_{\uparrow}, \: x_{\downarrow} \ne y_{\downarrow}$ into
account. The first attempt that can be made, is to consider
$x_{\uparrow} = x_{\downarrow} = x$, $y_{\uparrow} = y_{\downarrow} = y$, and
$x = y$. This solution presents the interesting property that reduces the 
system to 1D case. This means that the solution emerges only for $t^d_y =
t^f_y = V_y = 0$ and $t^d_{y \pm x} = t^f_{y \pm x} = V_{y \pm x} = 0$. New
aspects related to PAM in comparison with those reported in 
Refs.(\cite{exa11,exa12}) are not present. This case merits however attention
in the future, since it allows to study at the level of exact ground-states
(taken in the form of Eq.(\ref{e26})) the modification of the 1D properties
to 2D characteristics by taking into account small and smooth deviations from
the $x=y$ condition.

c) The third solution that we have found was deduced in 
$x_{\uparrow} = x_{\downarrow} = x$, $y_{\uparrow} = y_{\downarrow} = y$, and
$x \ne y$ case. This solution will be presented here in details, since presents
a 2D ground-state that emerges for zero next-nearest-neighbor $\hat H$
contributions. A such type of exact solution for PAM is completely new, 
because it cannot be obtained by the decomposition used previously
\cite{exa13}. 

d) We have studied also the general $x_{\sigma} \ne x_{-\sigma}, y_{\sigma'}$,
$y_{\sigma} \ne y_{-\sigma}, x_{\sigma'}$, case as well, obtaining only 
localized solution which require  the presence of next-nearest neighbor 
$\hat H$ terms as well.

\subsection{Detailed analysis of the solution obtained in the absence of
next-nearest neighbor Hamiltonian terms.}

Herewith, we analyze in detail the solution c) described above requiring
$x \ne y$. This emerges at
\begin{eqnarray}
t^d_{y \pm x} = t^f_{y \pm x} = V^{df}_{y \pm x} = V^{fd}_{y \pm x} = 0 \: ,
\label{e28}
\end{eqnarray}
so it describes a ground-state wave function for PAM not containing in its 
Hamiltonian NNN extension terms. A such type of exact ground-state in 2D
at finite nonzero value of the interaction is presented for the first time in
this paper.
  
Solving for the plaquette operator parameters the system of equations
Eqs.(\ref{d1},\ref{d2}) we have found
\begin{eqnarray}
&&a_{1,d,\uparrow} = x^{*} p_d \: , \quad a_{1,d,\downarrow} = \frac{
y x^{*}\tau_2^{*}}{z_1} p_d \: , \quad a_{2,d,\uparrow} = \frac{y^{*}}{
z_1^{*}} p_d \: , \quad a_{2,d,\downarrow} = - \frac{1}{\tau_1} p_d \: , 
\nonumber\\
&&a_{3,d,\uparrow} = \frac{\tau_2 y^{*}}{z_1^{*}} p_d \: , \quad
a_{3,d,\downarrow} = - p_d \: , \quad a_{4,d,\uparrow} = - \frac{1}{
y \tau_1^{*}} p_d \: , \quad a_{4,d,\downarrow} = - \frac{1}{z_1} p_d \: , 
\nonumber\\
&&a_{1,f,\uparrow} = x^{*} p_f \: , \quad a_{1,f,\downarrow} =
\frac{y x^{*} \tau_2^{*}}{z_1} p_f \: , \quad a_{2,f,\uparrow} = \frac{y^{*}}{
z_1^{*}} p_f \: , \quad a_{2,f,\downarrow} = - \frac{1}{\tau_1} p_f \: ,
\nonumber\\
&&a_{3,f,\uparrow} = \frac{\tau_2 y^{*}}{z_1^{*}} p_f \: , \quad
a_{3,f,\downarrow} = - p_f \: , \quad a_{4,f,\uparrow} = - \frac{1}{
y \tau_1^{*}} p_f \: , \quad a_{4,f,\downarrow} = - \frac{1}{z_1} p_f \: ,
\nonumber\\
&& b_{1,d,\uparrow} = p_d \: , \quad b_{1,d,\downarrow} = \frac{y \tau_2^{*}}{
z_1} p_d \: , \quad b_{2,d,\uparrow} = \frac{1}{z_1^{*}}p_d \: , \quad
b_{2,d,\downarrow} = - \frac{1}{\tau _1 y^{*}} p_d \: ,
\nonumber\\
&& b_{3,d,\uparrow} = - \frac{x \tau_2 y^{*}}{z_1^{*}} p_d \: , \quad
b_{3,d,\downarrow} = x p_d \: , \quad b_{4,d,\uparrow} = \frac{1}{\tau_1^{*}}
p_d \: , \quad b_{4,d,\downarrow} = \frac{y}{z_1} p_d \: ,
\nonumber\\
&&b_{1,f,\uparrow} = p_f \: , \quad b_{1,f,\downarrow} = \frac{y \tau_2^{*}}{
z_1} p_f \: , \quad b_{2,f,\uparrow} = \frac{1}{z_1^{*}} p_f \: , \quad
b_{2,f,\downarrow} = - \frac{1}{\tau_1 y^{*}} p_f \: ,
\nonumber\\
&&b_{3,f,\uparrow} = - \frac{x \tau_2 y^{*}}{z_1^{*}} p_f \: , \quad
b_{3,f,\downarrow} = x p_f \: , \quad b_{4,f,\uparrow} = \frac{1}{\tau_1^{*}}
p_f \: , \quad b_{4,f,\downarrow} = \frac{y}{z_1} p_f \: .
\label{e29}
\end{eqnarray}
The conditions imposed for the parameters entering in Eq.(\ref{e29}) are
$x y^{*} \ne -1$, $x \ne y$,  
$\tau_1 \ne \tau_2$, $\tau_1 \tau_2^{*}=$ real, $p_d p_f^{*} = $ real.
Together with Eq.(\ref{e29}), the nonzero $\hat H$ parameters become
\begin{eqnarray}
&&t^d_x = - R_1 |p_d|^2 \: , \quad t^d_y = -R_2 |p_d|^2 \: , \quad t^f_x = -
R_1 |p_f|^2 \: , \quad t^f_y = - R_2 |p_f|^2 \: ,
\nonumber\\
&&V_x = - R_1 p_d^{*} p_f \: , \quad V_y = - R_2 p_d^{*} p_f \: , \quad
V_0 = - R_3 p_d^{*} p_f \: , \quad U + E_f = K - K^f \: ,
\nonumber\\
&&K = R_3 |p_d|^2 \: , \quad K^f = R_3 |p_f|^2 \: .
\label{e30}
\end{eqnarray}
The $R_n$, $n=1,2,3$ factors present in this relation are given by
\begin{eqnarray}
&&R_1 = \frac{(1 + x y^{*})}{z_1^{*}} (1 - \frac{\tau_2}{\tau_1}) \: , \quad
R_2 = \frac{(y - x)}{y \tau_1^{*}} ( 1 + \frac{\tau_2 \tau_1^{*} |y|^2}{
|z_1|^2}) \: ,
\nonumber\\
&&R_3 = (1 + |x|^2) ( 1 + \frac{|\tau_2|^2 |y|^2}{|z_1|^2}) +
\frac{1 + |y|^2}{|z_1|^2} (1 + \frac{|z_1|^2}{|\tau_1|^2 |y|^2} ) \: .
\label{e31}
\end{eqnarray}
We further mention, that the obtained solution, for $y = z_1 =\tau_1$ and
$x = \tau_2$ reduces to the isotropic case, where $t^{g}_x = t^{g}_y =
t^{g}$, $g=d,f$, and $V_x = V_y = V$.

The ground-state wave function from Eq.(\ref{e26}) in the case of the solution
from Eq.(\ref{e29}) reduces to the simple form
\begin{eqnarray}
|\Psi_g(loc)\rangle = \prod_{\bf i} (p_d^{*} \hat d^{\dagger}_{{\bf i},
\uparrow} + p_f^{*} \hat f^{\dagger}_{{\bf i},\uparrow})(p_d^{*} \hat d^{
\dagger}_{{\bf i},\downarrow} + p_f^{*} \hat f^{\dagger}_{{\bf i},\downarrow})
(\mu_{{\bf i},\uparrow} \hat f^{\dagger}_{{\bf i},\uparrow} +
\mu_{{\bf i},\downarrow} \hat f^{\dagger}_{{\bf i},\downarrow})|0\rangle \: .
\label{e32}
\end{eqnarray}
The result presented in Eq.(\ref{e32}) is obtained because $\hat P_1 = 
\prod_{\bf i} \hat A_{\bf i}^{\dagger} \hat B_{\bf i}^{\dagger}$, in
the studied case, is unable to introduce three particles on the 
same lattice site. Since $\hat P_1$ introduces
$N=2N_{\Lambda}$ electrons in the system, being unable to put three electrons
on a given site, an uniform particle distribution will be obtained with two
electrons per site, which generates the product
$\prod_{\bf i} (p_d^{*} \hat d^{\dagger}_{{\bf i},
\uparrow} + p_f^{*} \hat f^{\dagger}_{{\bf i},\uparrow})(p_d^{*} \hat d^{
\dagger}_{{\bf i},\downarrow} + p_f^{*} \hat f^{\dagger}_{{\bf i},\downarrow})$
in $|\Psi_g(loc)\rangle$ in Eq.(\ref{e32}). The added term contained in
$\hat F_{\mu}$ (see Eq.(\ref{e26})) introduces one more $f$ electron on each 
site, and as a consequence, do not modify the created uniform 
particle distribution within the system, and Eq.(\ref{e32}) arise. This 
wave-function has a well defined norm
\begin{eqnarray}
\langle \Psi_g(loc)|\Psi_g(loc)\rangle = (|p_d|^2 + |p_f|^2)^{N_{\Lambda}}
\prod_{\bf i} (|\mu_{{\bf i},\uparrow}|^2 + |\mu_{{\bf i},\downarrow}|^2) \: ,
\label{e33}
\end{eqnarray}
and as mentioned above, coherently maintains three particles on every lattice 
site
\begin{eqnarray}
\hat n_{\bf i} |\Psi_g(loc)\rangle =
[\sum_{\sigma}(\hat d^{\dagger}_{{\bf i},\sigma} \hat d_{{\bf i},\sigma} +
\hat f^{\dagger}_{{\bf i},\sigma} \hat f_{{\bf i},\sigma} )]
|\Psi_g(loc)\rangle = 3 |\Psi_g(loc)\rangle \: .
\label{e34}
\end{eqnarray}
Denoting by $\langle ... \rangle =\langle \Psi_g(loc)| ... |\Psi_g(loc)\rangle
/ \langle \Psi_g(loc)|\Psi_g(loc)\rangle$ the ground-state expectation values,
we obtain long-range density-density correlations within the system
\begin{eqnarray}
\frac{1}{\langle \hat n_{\bf i} \rangle} \langle 
\hat n_{\bf i} \hat n_{{\bf j} \ne {\bf i}} \rangle = 3 \: .
\label{e35}
\end{eqnarray}
Furthermore, it can be observed that $|\Psi_g(loc)\rangle$ prohibits in the 
same time the hopping and non-local hybridization between all site pairs
\begin{eqnarray}
\langle \hat T_d \rangle = \langle \hat T_f \rangle =  
\langle \hat V_{nl} \rangle = 0 \: ,
\label{e36}
\end{eqnarray}
since, for $\hat g, \hat g' = \hat f, \hat d$, we have $\langle \hat g_{
{\bf i},\sigma}^{\dagger} \hat g'_{{\bf j},\sigma'} \rangle = 0$, for all
${\bf j} \ne {\bf i}$. As a consequence, the ground-state 
$|\Psi_g(loc)\rangle$ clearly represents a completely localized state.

The remaining non-zero ground-state expectation values of different $\hat H$
terms are given by
\begin{eqnarray}
\frac{\langle \hat E_f \rangle}{N_{\Lambda}} = E_f \frac{2|p_f|^2 + |p_d|^2}{
|p_d|^2 + |p_f|^2} \: , \quad 
\frac{\langle \hat U \rangle}{N_{\Lambda}} = U \frac{|p_f|^2 }{
|p_d|^2 + |p_f|^2} \: , \quad  
\frac{\langle \hat V_0 \rangle}{N_{\Lambda}} = - \frac{ K |p_f|^2 + 
K^f |p_d|^2}{|p_d|^2 + |p_f|^2} \: .
\label{e37}
\end{eqnarray}
Because of $U > 0$, and as seen from Eqs.(\ref{e30},\ref{e31}), $K,K^f > 0$, 
the non-zero on-site hybridization, coupling the two
bands, decreases the energy of the system ($\langle \hat V_0 \rangle < 0 , \:
\langle \hat E_f \rangle > 0 , \: \langle \hat U \rangle > 0$). The 
ground-state energy becomes $E_g = \langle \hat V_0 + \hat E_f + \hat U
\rangle$.

In average, the number of $f$-electrons per site becomes
\begin{eqnarray}
\langle \sum_{\sigma} \hat f^{\dagger}_{{\bf i},\sigma} \hat f_{{\bf i},
\sigma} \rangle = 1 + \frac{x^2}{1 + x^2} \: ,
\label{e38}
\end{eqnarray}
where, $x=p_f/p_d$ and we have $x^2=|t^f/t^d|$ in the isotropic case, and 
$x^2 = |t^f_{\alpha} / t^d_{\alpha}|$, $\alpha = x,y$ in general, for the
considered solution. Since in concrete physical 
situations $x << 1$, the number of $f$ electrons per site is close to one, but
not exactly one in the ground-state. Excepting the small number of sites with
double $f$-electron occupancy, the local $f$-moments are not compensated. In 
fact, defining $\hat m_{\bf i}^{g} = \hat n^{g}_{{\bf i},\uparrow} -
\hat n^{g}_{{\bf i},\downarrow}$, with $g=d,f$, we have
$\langle \hat m^d_{\bf i} \rangle = x \langle \hat m^f_{\bf i} \rangle$, and
$\langle \hat m^f_{\bf i} \rangle + \langle \hat m^d_{\bf i} \rangle =
(|\mu_{{\bf i},\uparrow}|^2 -|\mu_{{\bf i},\downarrow}|^2)/   
(|\mu_{{\bf i},\uparrow}|^2 +|\mu_{{\bf i},\downarrow}|^2)$, where as presented
before, $\mu_{{\bf i},\sigma}$ are arbitrary.

Concentrating on the magnetic properties of the ground-state, the
total spin of the system can be standardly expressed via
$S_z = 1/2 \sum_{\bf i} (\hat d^{\dagger}_{{\bf i},\uparrow}
\hat d_{{\bf i}, \uparrow} + \hat f^{\dagger}_{{\bf i},\uparrow}
\hat f_{{\bf i}, \uparrow} - \hat d^{\dagger}_{{\bf i},\downarrow}
\hat d_{{\bf i}, \downarrow} - \hat f^{\dagger}_{{\bf i},\downarrow}
\hat f_{{\bf i}, \downarrow})$, $S^{+} = \sum_{\bf i} (\hat d^{\dagger}_{
{\bf i},\uparrow} \hat d_{{\bf i}, \downarrow} + \hat f^{\dagger}_{{\bf i},
\uparrow} \hat f_{{\bf i}, \downarrow})$, $S^{-} = (S^{+})^{\dagger}$, and
$S^2 = S_z^2 + 1/2(S^{+} S^{-} + S^{-} S^{+})$. Calculating the ground-state
expectation values, we find
\begin{eqnarray}
&&\langle S^2 \rangle = \frac{3 N_{\Lambda}}{4} + \sum_{{\bf i} \ne {\bf j}}
\frac{(|\mu_{{\bf i},\uparrow}|^2 -|\mu_{{\bf i},\downarrow}|^2)   
(|\mu_{{\bf j},\uparrow}|^2 - |\mu_{{\bf j},\downarrow}|^2) +
2 (\mu_{{\bf i},\downarrow} \mu^{*}_{{\bf i},\uparrow}   
\mu_{{\bf j},\uparrow} \mu^{*}_{{\bf j},\downarrow} + c.c.)}{
4(|\mu_{{\bf i},\uparrow}|^2 +|\mu_{{\bf i},\downarrow}|^2)   
(|\mu_{{\bf j},\uparrow}|^2 +|\mu_{{\bf j},\downarrow}|^2)} \: ,
\nonumber\\
&&\langle S^2_z \rangle = \frac{N_{\Lambda}}{4} + \sum_{{\bf i} \ne {\bf j}}
\frac{(|\mu_{{\bf i},\uparrow}|^2 -|\mu_{{\bf i},\downarrow}|^2)   
(|\mu_{{\bf j},\uparrow}|^2 - |\mu_{{\bf j},\downarrow}|^2) }{
4(|\mu_{{\bf i},\uparrow}|^2 +|\mu_{{\bf i},\downarrow}|^2)   
(|\mu_{{\bf j},\uparrow}|^2 +|\mu_{{\bf j},\downarrow}|^2)} \: .
\label{e39}
\end{eqnarray}
Taking now two extremum $\{\mu_{{\bf i},\sigma} \}$ distributions, 
a) for $\mu_{{\bf i},\uparrow} = \mu_{\uparrow}, \: \mu_{{\bf i},\downarrow} 
= 0$,
we find $\langle S_z^2 \rangle = (N_{\Lambda}/2)^2$, and  $\langle S^2 \rangle
= (N_{\Lambda}/2)(N_{\Lambda}/2+1)$. This situation corresponds to maximum 
total spin in the system, with average total spin absolute value per site
$(\langle S^2 \rangle / N_{\Lambda}^2)^{1/2} = \sqrt{1/2(1/2 + 
1/N_{\Lambda})}$, which is of order $1/2$ for large $N_{\Lambda}$. b) Dividing
however the square lattice into two equal sub-lattices with 
$\mu_{{\bf i},\uparrow} = \mu, \: \mu_{{\bf i},\downarrow} = 0$
in one sub-lattice, and $\mu_{{\bf i},\uparrow} = 0, \: \mu_{{\bf i},\downarrow}
= \mu$ in the other one, we obtain $\langle S_z^2 \rangle = 0 , \:
\sqrt{\langle S^2 \rangle / N^2_{\Lambda}} = 1/(\sqrt{2 N_{\Lambda}})$, i.e. 
in the thermodynamic limit, the total spin in absolute value per site is zero 
in this case.
As can be observed, the degeneration of the ground-state physically is given 
by the fact that all possible total spin values are contributing in its
construction. As a consequence, the ground-state behaves paramagnetically.
We must further observe, that not all different  $\{\mu_{{\bf i},\sigma} \}$
sets provide linearly independent ground-state wave-function contributions. 
For example,
choosing for all ${\bf i}$ the values $\mu_{{\bf i},\uparrow} = \mu_{\uparrow},
\: \mu_{{\bf i},\downarrow} = 0$, or  $\mu_{{\bf i},\uparrow} = 0,
\: \mu_{{\bf i},\downarrow} = \mu_{\downarrow}$, or $\mu_{{\bf i},\uparrow} = 
\mu_{\uparrow}, \: \mu_{{\bf i},\downarrow} = \mu_{\downarrow}$, we recover the
same ground-state with maximum value of $\langle S^2 \rangle $. As a 
consequence, in order to find orthogonal wave-functions that belong to the
ground-state, the $\mu_{{\bf i},\sigma}$ coefficients cannot be chosen
completely random and independent. Also the normalization to unity of 
$|\Psi_g(loc)\rangle$ 
represents a constraint to the value of these coefficients. Neglecting the 
trivial multiplicity obtained from the spatial orientation of the total spin 
${\vec S}$, the degree of the degeneration of the ground-state is 
$N_{\Lambda}/2$.

The spin-spin correlation functions can be also calculated, and for
${\bf i} \ne {\bf j}$ we obtain in the case of a fixed 
$\{ \mu_{{\bf i},\sigma} \}$ set 
\begin{eqnarray}
\langle {\vec S}_{\bf i} \cdot {\vec S}_{\bf j} \rangle = \frac{1}{4}
\frac{(|\mu_{{\bf i},\uparrow}|^2 -|\mu_{{\bf i},\downarrow}|^2)   
(|\mu_{{\bf j},\uparrow}|^2 - |\mu_{{\bf j},\downarrow}|^2) +
2 (\mu_{{\bf i},\uparrow} \mu^{*}_{{\bf i},\downarrow}   
\mu^{*}_{{\bf j},\uparrow} \mu_{{\bf j},\downarrow} + c.c.)}{
(|\mu_{{\bf i},\uparrow}|^2 +|\mu_{{\bf i},\downarrow}|^2)   
(|\mu_{{\bf j},\uparrow}|^2 +|\mu_{{\bf j},\downarrow}|^2)} \: .
\label{e40}
\end{eqnarray}
Since, as shown before, the $\mu_{{\bf i},\sigma}$ coefficients are not 
completely independent, the spin-spin correlations given by Eq.(\ref{e40}) are
quasi-random. Resembling behavior is experimentally seen in heavy-fermion 
cases \cite{x0}.

The phase diagram region where the solution occurs is presented in Fig. 5. for
the isotropic case. The general aspect of this region remains the same in the
anisotropic case as well. It represents a surface in the parameter space which 
extends from the low $U$ region up to the high $U$ region as well. This region
is completely different from that obtained in Ref.(\cite{exa13}) which emerges
for nonzero values of next-nearest-neighbor hopping and non-local 
hybridizations. 

The non-local nearest-neighbor hybridization matrix element in the isotropic 
case is related to hopping matrix elements by $(V/t^d)^2 = t^f/t^d$. In the 
anisotropic case this relation becomes $(V_{\alpha}/t^d_{\alpha})^2 =
t^f_{\alpha}/t^d_{\alpha}$, $\alpha = x,y$. Modifying the values of hopping
or/and hybridization matrix elements, we can leave ${\cal P}_H$, destroying
the ground-state character of $|\Psi_g(loc)\rangle$. This process can be tuned
by pressure which strongly influences the $t_{\bf r}, V_{\bf r}$ parameters
(see for example Ref.(\cite{xp})).
Since the reduction of Eq.(\ref{e26}) into the completely localised 
$|\Psi_g(loc)\rangle$ from Eq.(\ref{e32}) it is itself based on a delicate
balance between $\hat H$ parameters (contained in Eqs.(\ref{e30})), the loss of
the localization character of particles in principle can be easily achieved.
This localization-delocalization transition represents in fact a MIT transition
provided by PAM. Since the exactly deduced ground-state energy do not contains
the exponential term characteristic for a Kondo type behavior (see for example
the discussion presented in Ref.(\cite{int2})), a such type of MIT transition 
cannot be connected to Kondo physics. Instead, the MIT transition
connected to the destruction of the localized $|\Psi_g(loc)\rangle$ 
ground-state is related to the break-down of the long-range density-density 
correlations.

\section{Summary and conclusions}

We are presenting for the first time exact ground-states for the periodic
Anderson model (PAM) at finite $U$ in $D=2$ dimensions in the case in which the
Hamiltonian does not contain next to nearest neighbor (NNN) extension terms. 
For this reason, and based on this frame 
(i) The used plaquette operator procedure is presented in detail
and it is underlined that its applicability is not connected to the presence
of NNN extension terms in the Hamiltonian. We
underline that this is the only one procedure known at the moment, which is 
able to provide exact ground-states for PAM in $D=2$ dimensions and finite 
$U$ interaction values.
(ii) A new plaquette operator has been introduced for the study of the PAM
Hamiltonian. The plaquette operator contains contributions coming from all
spin components, possesses spin dependent numerical prefactors, and allows the
detection of ground-states even in the absence of NNN extension terms in the
Hamiltonian in restricted regions of the parameter space. (iii) The physical 
properties of the deduced ground-state have been analysed in detail. All 
relevant ground-state expectation values and correlation functions have been 
deduced for this reason. (iv) The implications of the deduced results relating
the metal-insulator transition in the frame of PAM have been analysed. 
It has been pointed out that the lost of the localization character in the
studied case is connected to the break-down of the long-range density-density
correlations rather than Kondo physics.  

The obtained new exact ground-state emerges at $3/4$ filling, not requires
the presence of NNN extension terms in the Hamiltonian, it is paramagnetic, 
and presents quasi-random spin-spin correlations. It represents a fully 
quantum-mechanical state (in the sense that it is far to be quasi-classical), 
and it is build up through superposition effects. The ground-state 
wave-function coherently controls the occupation number on all lattice sites, 
introducing in this manner long-range density-density correlations within the 
system, and prohibiting in the same time the hopping and non-local 
hybridizations. The local $f$ moments are not compensated and the $f$ electron
occupation number per site in average is close to, but not exactly one.  

Concerning the question of the physical relevance, we would like to mention
that in general terms, even a solution detected in a restricted 
parameter-space region which behaves completely repulsively in the 
renormalization group language could have significant physical implications 
\cite{x1}. In the present case, besides presenting open roots toward the
deduction possibilities of exact ground-states in $D=2$ dimensions for strongly
correlated systems, the presented results provide exact theoretical 
data which can be used in the process of understanding and description of the 
metal-insulator transition in the frame of PAM. 

\acknowledgements

The research has been supported in 2002 by the contract OTKA-T-037212 of 
Hungarian founds for scientific research. The author kindly acknowledge 
extremely valuable discussions on the subject with Dieter Vollhardt. 
He also would like to thank for the kind hospitality of the Department 
of Theoretical Physics III., University Augsburg in autumn 2001, 4 months of  
working period relating this field spent there, and supported by Alexander 
von Humboldt Foundation. 

\newpage
\appendix

\section{The explicit expression of the non-local one particle contributions
in the Hamiltonian.}

The kinetic energy term for $d$ electrons has the explicit form
\begin{eqnarray}
\hat T_d &=& \sum_{{\bf i},\sigma} [ (t^d_x \hat d^{\dagger}_{{\bf i},\sigma}
\hat d_{{\bf i} + {\bf x},\sigma} + H.c. ) + (t^d_y \hat d^{\dagger}_{{\bf i},
\sigma} \hat d_{{\bf i} + {\bf y},\sigma} + H.c. ) 
\nonumber\\
&+& (t^d_{x+y} \hat d^{\dagger}_{{\bf i},\sigma} \hat d_{{\bf i} + ({\bf x} +
{\bf y}),\sigma} + H.c. ) + (t^d_{y-x} \hat d^{\dagger}_{{\bf i},
\sigma} \hat d_{{\bf i} + ({\bf y}-{\bf x}),\sigma} + H.c. ) ] \: .
\label{a1}
\end{eqnarray}
The kinetic energy term for $f$ electrons can be simply obtained from Eq.(
\ref{a1}) interchanging the $d$ index with the $f$ index, and $\hat d$ by 
$\hat f$.

The explicit expression of the non-local hybridization becomes
\begin{eqnarray}
\hat V_{nl} &=& \sum_{{\bf i},\sigma} [
(V^{df}_{x} \hat d^{\dagger}_{{\bf i},\sigma} \hat f_{{\bf i}+{\bf x},\sigma}
+ H.c. ) +
(V^{fd}_{x} \hat f^{\dagger}_{{\bf i},\sigma} \hat d_{{\bf i}+{\bf x},\sigma}
+ H.c. ) 
\nonumber\\
&+&
(V^{df}_{y} \hat d^{\dagger}_{{\bf i},\sigma} \hat f_{{\bf i}+{\bf y},\sigma}
+ H.c. ) +
(V^{fd}_{y} \hat f^{\dagger}_{{\bf i},\sigma} \hat d_{{\bf i}+{\bf y},\sigma}
+ H.c. ) 
\nonumber\\
&+&
(V^{df}_{x+y} \hat d^{\dagger}_{{\bf i},\sigma} \hat f_{{\bf i}+({\bf x}+
{\bf y}),\sigma} + H.c. ) +
(V^{fd}_{x+y} \hat f^{\dagger}_{{\bf i},\sigma} \hat d_{{\bf i}+({\bf x}+
{\bf y}),\sigma} + H.c. ) 
\nonumber\\
&+&
(V^{df}_{y-x} \hat d^{\dagger}_{{\bf i},\sigma} \hat f_{{\bf i}+({\bf y}-
{\bf x}),\sigma} + H.c. ) +
(V^{fd}_{y-x} \hat f^{\dagger}_{{\bf i},\sigma} \hat d_{{\bf i}+({\bf y}-
{\bf x}),\sigma} + H.c. ) ]  \: .
\label{a2}
\end{eqnarray}

\section{The plaquette operator contributions summed up over the lattice 
sites.}

The expression $\hat A^{\dagger}_{\bf i} \hat A_{\bf i}$ summed up over the 
whole lattice considered with periodic boundary conditions in both directions
is presented below in condensed form ($\hat g, \hat g' = \hat d, \hat f$ ;
$g,g'=d,f$).
\begin{eqnarray}
&&\sum_{\bf i} \hat A^{\dagger}_{\bf i} \hat A_{\bf i} =
\sum_{\sigma,\sigma'} \sum_{g,g'} \sum_{\bf i}^{N_{\Lambda}} \{
\nonumber\\
&&[\hat g^{\dagger}_{{\bf i},\sigma} \hat g'_{{\bf i} + {\bf x}, \sigma'}
(a^{*}_{1,g,\sigma} a_{2,g',\sigma'} + a^{*}_{4,g,\sigma} a_{3,g',\sigma'})
+ H.c.] +
[\hat g^{\dagger}_{{\bf i},\sigma} \hat g'_{{\bf i} + {\bf y}, \sigma'}
(a^{*}_{1,g,\sigma} a_{4,g',\sigma'} + a^{*}_{2,g,\sigma} 
a_{3,g',\sigma'}) + H.c.] +
\nonumber\\
&&[\hat g^{\dagger}_{{\bf i},\sigma} \hat g'_{{\bf i} + ({\bf x}+{\bf y}), 
\sigma'} (a^{*}_{1,g,\sigma} a_{3,g',\sigma'}) + H.c.] +
[\hat g^{\dagger}_{{\bf i},\sigma} \hat g'_{{\bf i} + ({\bf y}-{\bf x}), 
\sigma'} (a^{*}_{2,g,\sigma} a_{4,g',\sigma'}) + H.c.] +
\nonumber\\
&&[\hat g^{\dagger}_{{\bf i},\sigma} \hat g'_{{\bf i}, \sigma'} 
(\sum_{n=1}^4 a^{*}_{n,g,\sigma} a_{n,g',\sigma'}) + H.c.] [ 1 -
\frac{1}{2} \delta_{g,g'} \delta_{\sigma,\sigma'} ] \} \: .
\label{b1}
\end{eqnarray}

Furthermore, the following property is satisfied
\begin{eqnarray}
\hat A^{\dagger}_{\bf i} \hat A_{\bf i} + \hat A_{\bf i} \hat A^{\dagger}_{
\bf i} = \sum_{n=1}^4 (|a_{n,d,\uparrow}|^2 + |a_{n,d,\downarrow}|^2 +
|a_{n,f,\uparrow}|^2 + |a_{n,f,\downarrow}|^2 ) \: .
\label{b2}
\end{eqnarray}

For the $\hat B_{\bf i}$ plaquette operators the Eqs.(\ref{b1},\ref{b2}) hold
as well by changing the coefficients $a_{n,g,\sigma}$ to $b_{n,g,\sigma}$,
where $g = d,f$.

\section{The nonlinear system of equations.}

The explicit expression of the system of equations Eq.(\ref{e16}) containing
70 equalities is presented
below in condensed form. The used abreviations are $g,g'=d,f$, 
$\sigma=\uparrow, \downarrow$, and $(g,g_1)$ represents $(d,f)$ or $(f,d)$. 
\begin{eqnarray}
- t^g_x &=& a^{*}_{1,g,\sigma} a_{2,g,\sigma} +
a^{*}_{4,g,\sigma} a_{3,g,\sigma} +  b^{*}_{1,g,\sigma} b_{2,g,\sigma}
+ b^{*}_{4,g,\sigma} b_{3,g,\sigma} \: , 
\nonumber\\
- t^g_y &=& a^{*}_{1,g,\sigma} a_{4,g,\sigma} +
a^{*}_{2,g,\sigma} a_{3,g,\sigma} +  b^{*}_{1,g,\sigma} b_{4,g,\sigma}
+ b^{*}_{2,g,\sigma} b_{3,g,\sigma} \: , 
\nonumber\\
0 &=&  a^{*}_{1,g,\sigma} a_{2,g',-\sigma} + a^{*}_{4,g,\sigma} 
a_{3,g',-\sigma} + b^{*}_{1,g,\sigma} b_{2,g',-\sigma} + 
b^{*}_{4,g,\sigma} b_{3,g',-\sigma} \: ,
\nonumber\\
0 &=&  a^{*}_{1,g,\sigma} a_{4,g',-\sigma} + a^{*}_{2,g,\sigma} 
a_{3,g',-\sigma} + b^{*}_{1,g,\sigma} b_{4,g',-\sigma} + 
b^{*}_{2,g,\sigma} b_{3,g',-\sigma} \: ,
\nonumber\\
- t^g_{x+y} &=&  a^{*}_{1,g,\sigma} a_{3,g,\sigma} + 
b^{*}_{1,g,\sigma} b_{3,g,\sigma} \: , \quad 
- t^g_{y-x} =  a^{*}_{2,g,\sigma} a_{4,g,\sigma} + 
b^{*}_{2,g,\sigma} b_{4,g,\sigma} \: , 
\nonumber\\
0 &=& a^{*}_{1,g,\sigma} a_{3,g',-\sigma} + 
b^{*}_{1,g,\sigma} b_{3,g',-\sigma} \: , \quad
0 = a^{*}_{2,g,\sigma} a_{4,g',-\sigma} + 
b^{*}_{2,g,\sigma} b_{4,g',-\sigma} \: ,
\nonumber\\
- V^{g,g_1}_x &=& a^{*}_{1,g,\sigma} a_{2,g_1,\sigma} +
a^{*}_{4,g,\sigma} a_{3,g_1,\sigma} +  b^{*}_{1,g,\sigma} b_{2,g_1,\sigma}
+ b^{*}_{4,g,\sigma} b_{3,g_1,\sigma} \: , 
\nonumber\\
- V^{g,g_1}_y &=& a^{*}_{1,g,\sigma} a_{4,g_1,\sigma} +
a^{*}_{2,g,\sigma} a_{3,g_1,\sigma} +  b^{*}_{1,g,\sigma} b_{4,g_1,\sigma}
+ b^{*}_{2,g,\sigma} b_{3,g_1,\sigma} \: , 
\nonumber\\
- V^{g,g_1}_{x+y} &=&  a^{*}_{1,g,\sigma} a_{3,g_1,\sigma} + 
b^{*}_{1,g,\sigma} b_{3,g_1,\sigma} \: , \quad 
- V^{g,g_1}_{y-x} =  a^{*}_{2,g,\sigma} a_{4,g_1,\sigma} + 
b^{*}_{2,g,\sigma} b_{4,g_1,\sigma} \: , 
\nonumber\\
- V_0 &=& \sum_{n=1}^4 a^{*}_{n,d,\sigma} a_{n,f,\sigma} + 
\sum_{n=1}^4 b^{*}_{n,d,\sigma} b_{n,f,\sigma} \: ,  
\nonumber\\
0 &=& \sum_{n=1}^4 a^{*}_{n,g,\sigma} a_{n,g',-\sigma} + 
\sum_{n=1}^4 b^{*}_{n,g,\sigma} b_{n,g',-\sigma} \: . 
\label{c1}
\end{eqnarray}

\section{The equations for the plaquette operator parameters}

After using Eq.(\ref{e27}), the remaining equations for the plaquette operator 
parameters are presented in detail in this Appendix. These equations can be 
divided in two parts: a homogeneous part (Eq.(\ref{d1})), and a 
non-homogeneous one (Eq.(\ref{d2})), see below. These two system of equations 
are presented here in detail in condensed form. Written explicitely, 
Eq.(\ref{d1}) (Eq.(\ref{d2})) contains 20 (41) different equations, 
respectively.

The homogeneous part of the equations is as follows ($\sigma=\uparrow,
\downarrow$, $g=d,f$, $g'=d,f$)
\begin{eqnarray}
&&(x_{\sigma} x^{*}_{-\sigma} + 1) b^{*}_{1,g,\sigma} b_{1,g',-\sigma}
+(y_{\sigma} y^{*}_{-\sigma} + 1) b^{*}_{2,g,\sigma} b_{2,g',-\sigma} +
\nonumber\\
&&(\frac{1}{x_{\sigma} x^{*}_{-\sigma}} + 1) b^{*}_{3,g,\sigma} 
b_{3,g',-\sigma} +
(\frac{1}{y_{\sigma} y^{*}_{-\sigma}} + 1) b^{*}_{4,g,\sigma} 
b_{4,g',-\sigma} = 0 \: ,
\nonumber\\ 
&&(x_{\sigma} y^{*}_{-\sigma} + 1) b^{*}_{1,g,\sigma} b_{2,g',-\sigma}
+(\frac{1}{x_{\sigma} y^{*}_{-\sigma}} + 1) b^{*}_{4,g,\sigma} 
b_{3,g',-\sigma} = 0 \: ,
\nonumber\\
&&(- \frac{x_{\sigma}}{ y_{\sigma}} + 1) b^{*}_{1,g,\sigma} 
b_{4,g',-\sigma} + (- \frac{y_{\sigma}}{x_{\sigma}} + 1) 
b^{*}_{2,g,\sigma} b_{3,g',-\sigma} = 0 \: ,
\label{d1}
\end{eqnarray}

The non-homogeneous part of the equations is presented below. The abbreviations
used here are $(g,g_1) =$ $(d,f)$ or $(f,d)$, and the presence of $\sigma$ 
means that two equations are simultaneously present with $\sigma=\uparrow$ and
$\sigma=\downarrow$. For a single $g$ index we have $g=d,f$.
\begin{eqnarray}
&&-V_0 =
(|x_{\sigma}|^2+1)b^{*}_{1,d,\sigma} b_{1,f,\sigma} +
(|y_{\sigma}|^2+1)b^{*}_{2,d,\sigma} b_{2,f,\sigma} +
\nonumber\\
&&(\frac{1}{|x_{-\sigma}|^2}+1)b^{*}_{3,d,\sigma} b_{3,f,\sigma} +
(\frac{1}{|y_{-\sigma}|^2}+1)b^{*}_{4,d,\sigma} b_{4,f,\sigma} \: ,
\nonumber\\
&&-t^g_x =
(x_{\sigma} y^{*}_{\sigma} + 1)b^{*}_{1,g,\sigma} b_{2,g,\sigma} +
(\frac{1}{x_{-\sigma} y^{*}_{-\sigma}} + 1)b^{*}_{4,g,\sigma} 
b_{3,g,\sigma} \: ,
\nonumber\\
&&-t^g_y =
(-\frac{x_{\sigma}}{y_{-\sigma}} + 1)b^{*}_{1,g,\sigma} b_{4,g,
\sigma} + (-\frac{y_{\sigma}}{x_{-\sigma}} + 1)b^{*}_{2,g,\sigma} 
b_{3,d,\sigma} \: ,
\nonumber\\
&&-V^{g,g_1}_x =
(x_{\sigma} y^{*}_{\sigma} + 1)b^{*}_{1,g,\sigma} b_{2,g_1,\sigma} +
(\frac{1}{x_{-\sigma} y^{*}_{-\sigma}} + 1)b^{*}_{4,g,\sigma} 
b_{3,g_1,\sigma} \: ,
\nonumber\\
&&-V^{g,g_1}_y =
(-\frac{x_{\sigma}}{y_{-\sigma}} + 1)b^{*}_{1,g,\sigma} b_{4,g_1,
\sigma} + (-\frac{y_{\sigma}}{x_{-\sigma}} + 1)b^{*}_{2,g,\sigma} 
b_{3,g_1,\sigma} \: ,
\nonumber\\
&&K^g = (|x_{\sigma}|^2+1)|b_{1,g,\sigma}|^2 +
(|y_{\sigma}|^2+1)|b_{2,g,\sigma}|^2 +
\nonumber\\
&&(\frac{1}{|x_{-\sigma}|^2} + 1)|b_{3,g,\sigma}|^2 +
(\frac{1}{|y_{-\sigma}|^2} + 1)|b_{4,g,\sigma}|^2 \: ,
\nonumber\\
&&-t^g_{x+y} = (-\frac{x_{\sigma}}{x_{-\sigma}} + 1)
b^{*}_{1,g,\sigma} b_{3,g,\sigma} \: , \quad
-t^g_{y-x} = (-\frac{y_{\sigma}}{y_{-\sigma}} + 1)
b^{*}_{2,g,\sigma} b_{4,g,\sigma} \: ,
\nonumber\\
&&-V^{g,g_1}_{x+y} = (-\frac{x_{\sigma}}{x_{-\sigma}} + 1)
b^{*}_{1,g,\sigma} b_{3,g_1,\sigma} \: , \quad
-V^{g,g_1}_{y-x} = (-\frac{y_{\sigma}}{y_{-\sigma}} + 1)
b^{*}_{2,g,\sigma} b_{4,g_1,\sigma} \: ,
\nonumber\\
&&V^{df}_{\alpha} = V^{fd}_{\alpha} = V_{\alpha} \: , \quad
\alpha = x, y, x + y, y - x \: ; \quad
U + E_f = K - K^f \: , K = K^d \: . 
\label{d2}
\end{eqnarray}

\newpage

\begin{figure}[h]
\centerline{\epsfbox{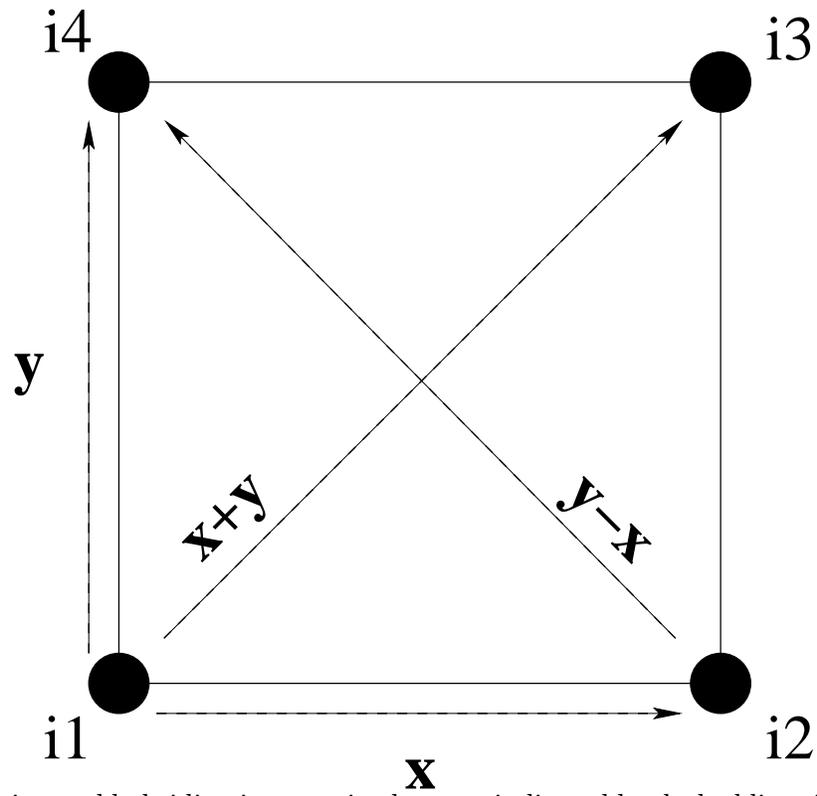}}
\caption{Hopping and hybridization matrix elements indicated by dashed lines
in the elementary plaquette $(i1,i2,i3,i4)$.}
\label{fig1}
\end{figure}

\newpage

\begin{figure}[h]
\centerline{\epsfbox{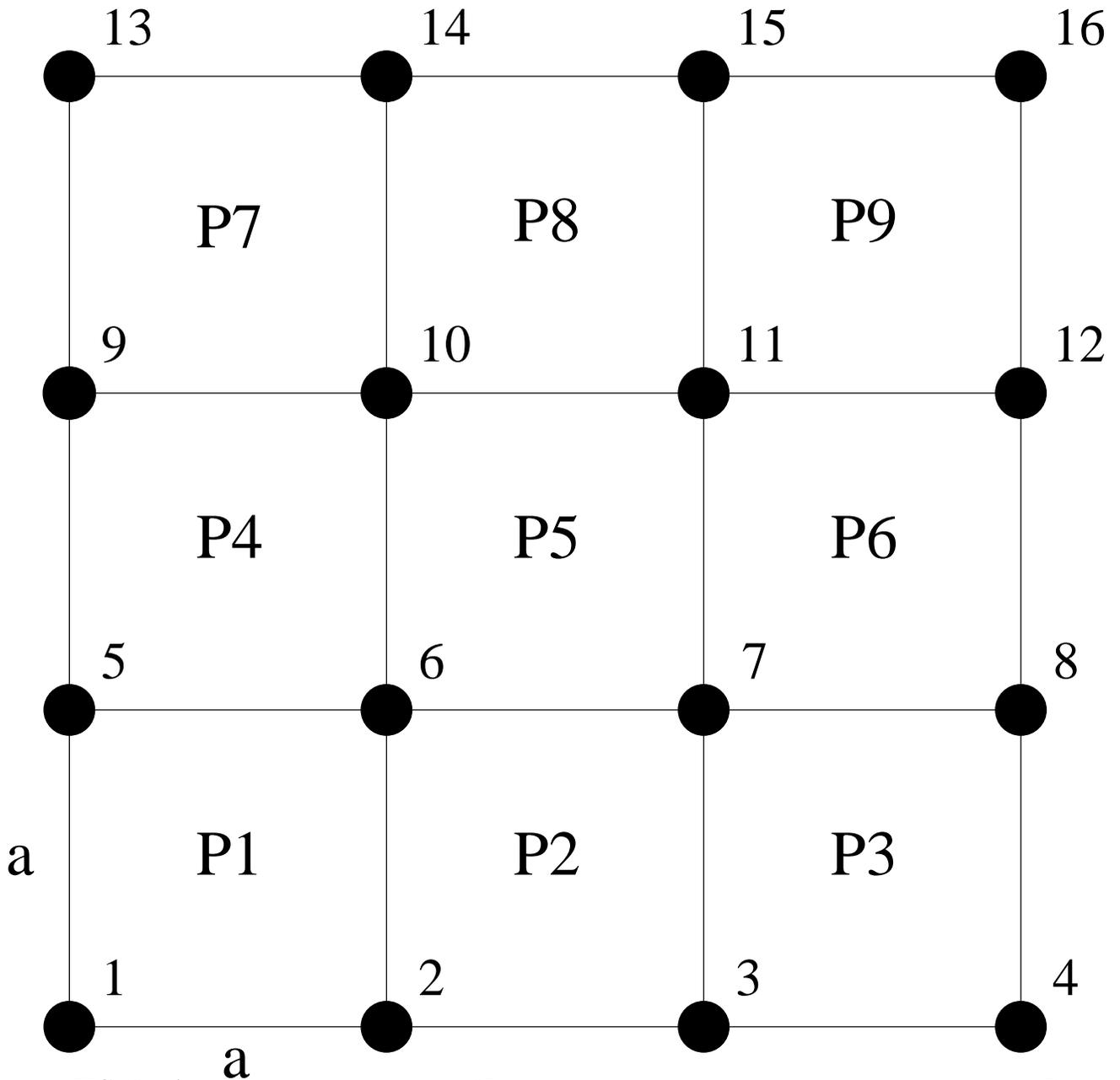}}
\caption{A $4 \times 4$ square lattice in 2D covered by elementary plaquettes
denoted by $PI$, $I=1,2,...9$. The numbers paced in down-left corner of every
plaquette denote the lattice sites, and $a$ is the lattice constant.}
\label{fig2}
\end{figure}

\newpage

\begin{figure}[h]
\centerline{\epsfbox{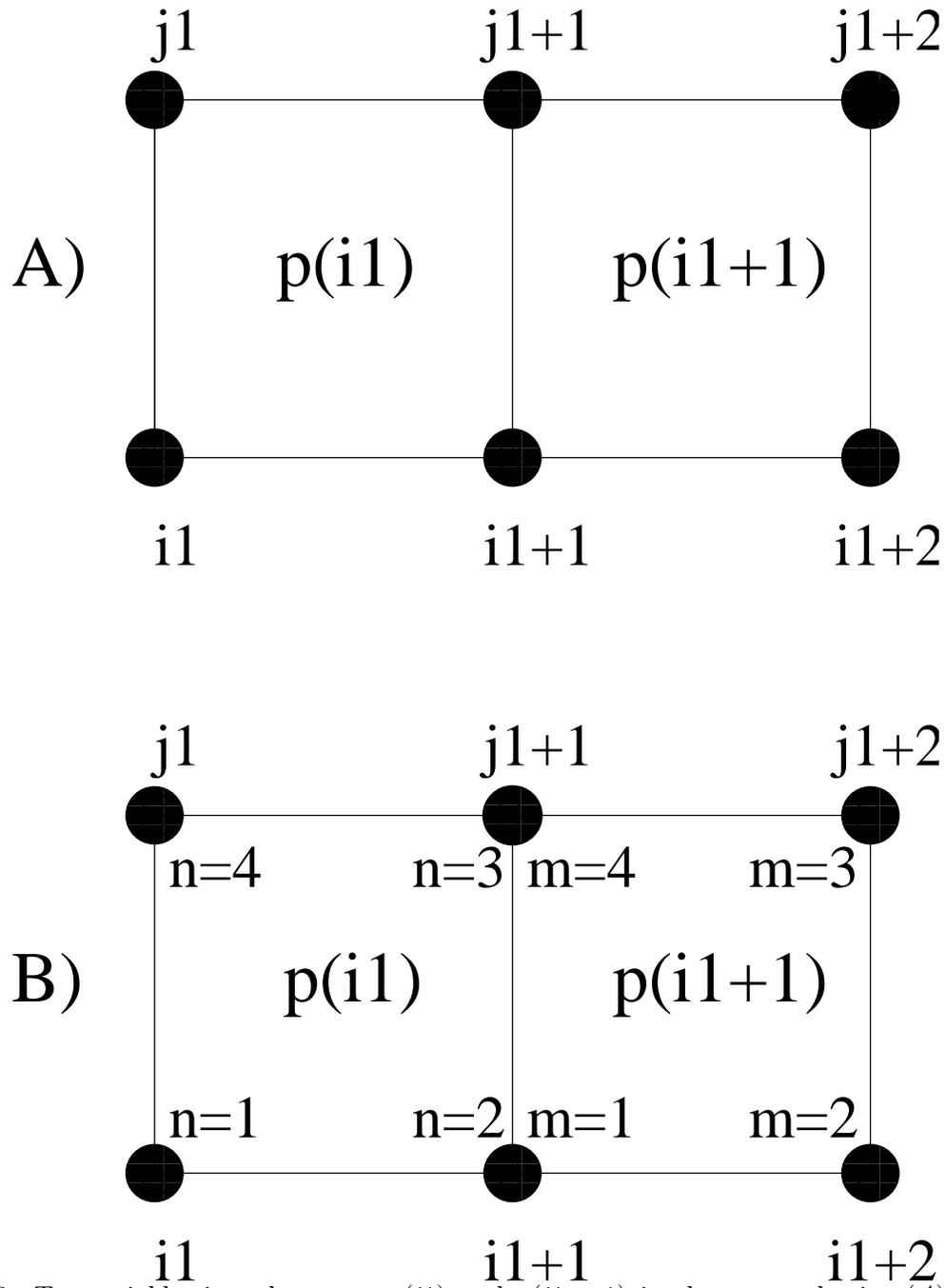}}
\caption{Two neighboring plaquettes $p(i1)$ and $p(i1+1)$ in the square 
lattice $(A)$, and the notation of lattice sites inside the plaquettes by $n$
and $m$ respectively $(B)$.}
\label{fig3}
\end{figure}

\newpage

\begin{figure}[h]
\centerline{\epsfbox{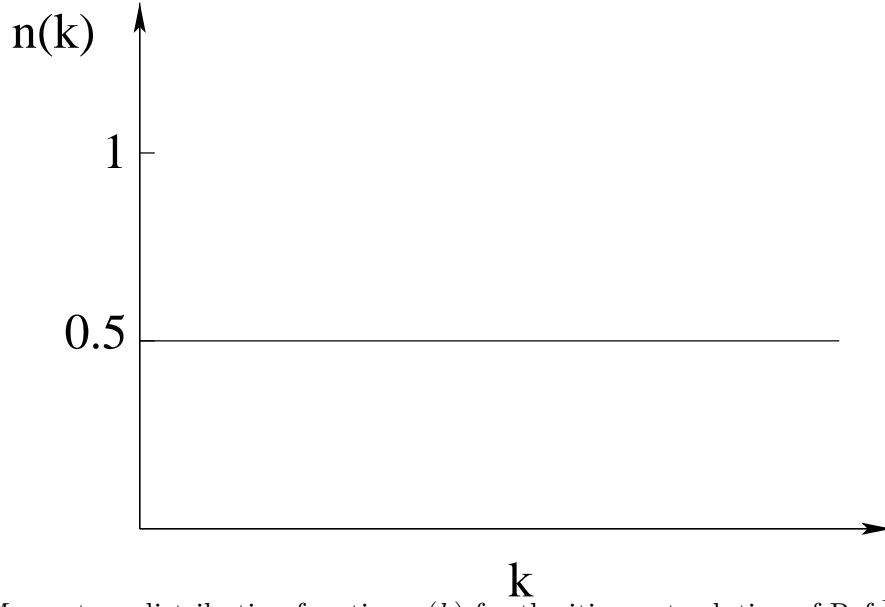}}
\caption{Momentum distribution function $n(k)$ for the itinerant solution of 
Ref.\cite{exa13} in the upper diagonalized, half filled band, along the whole
first Brillouin zone. As can be seen, neregularities of any kind in $n(k)$ and
its derivatives of any order are missing, signalling non-Fermi liquid behavior 
in normal phase and 2D.}
\label{fig4}
\end{figure}

\newpage

\begin{figure}[h]
\centerline{\epsfbox{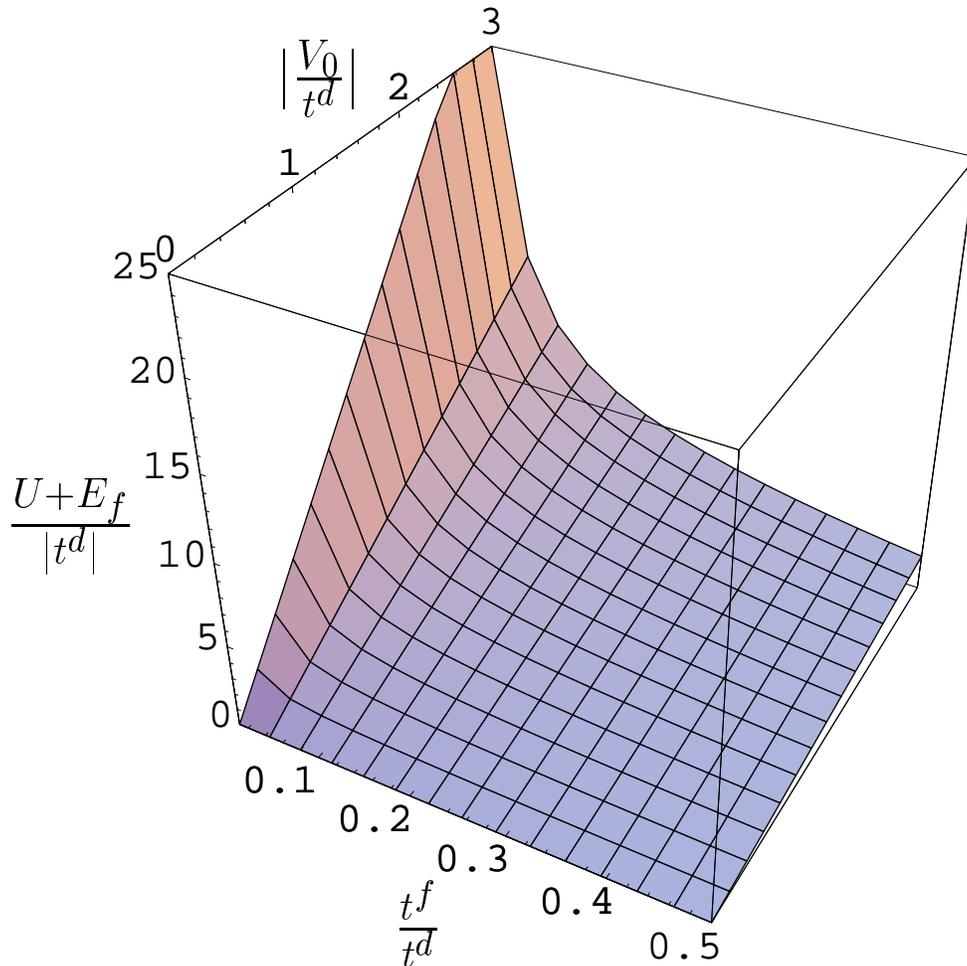}}
\caption{Phase diagram region where the localized solution occurs in the 
absence of next-nearest neighbor terms and isotropic case. The presented
surface extends up to infinity for $(U+E_f)/|t^d| \to \infty$.}
\label{fig5}
\end{figure}

\end{document}